\documentclass[sn-mathphys-num]{sn-jnl}

\usepackage{latexsym,epsfig,graphicx,amsmath,amssymb,amscd,multirow,paralist,dsfont,indentfirst,bm, adjustbox, color, colortbl, float, subfig, booktabs, url, algorithm, algorithmic,appendix, rotating, tikz, xurl}

\providecommand{\keywords}[1]{\textbf{\textit{keywords:}} #1}
\RequirePackage{amsthm,amsmath,amsfonts,amssymb}

\makeatletter
\def\@author#1{\g@addto@macro\elsauthors{\normalsize%
    \def\baselinestretch{1}%
    \upshape\authorsep#1\unskip\textsuperscript{%
     \ifx\@fnmark\@empty\else\unskip\sep\@fnmark\let\sep=,\fi
      \ifx\@corref\@empty\else\unskip\sep\@corref\let\sep=,\fi
      }%
    \def\authorsep{\unskip,\space}%
    \global\let\@fnmark\@empty
    \global\let\@corref\@empty  
    \global\let\sep\@empty}%
    \@eadauthor={#1}
}
\makeatother

\begin{document}


        

\title [Analyzing Taiwanese traffic patterns on consecutive holidays through forecast reconciliation and prediction-based anomaly detection techniques]{Analyzing Taiwanese traffic patterns on consecutive holidays through forecast reconciliation and prediction-based anomaly detection techniques}

\author[1]{\fnm{Mahsa} \sur{Ashouri}}\email{mashouri@umich.edu}

\author[2]{\fnm{Frederick Kin Hing} \sur{Phoa}}\email{fredphoa@stat.sinica.edu.tw}

\author*[3,4]{\fnm{Marzia A.} \sur{Cremona}}\email{marzia.cremona@fsa.ulaval.ca}

\affil[1]{\orgdiv{Department of Biostatistics}, \orgname{University of Michigan}, 
\country{US}}

\affil[2]{\orgdiv{Institute of Statistical Science}, \orgname{Academia Sinica}, 
\city{Taipei}, 
\country{Taiwan}}

\affil[3]{\orgdiv{Department of Operations and Decision Systems}, \orgname{Universit\'e Laval}, 
\city{Qu\'ebec}, 
\country{Canada}}
\affil[4]{\orgname{Interuniversity Research Center on Enterprise Networks, Logistics, and Transportation (CIRRELT)}, 
\country{Canada}}

\abstract{
This study explores traffic patterns on Taiwanese highways during consecutive holidays and focuses on understanding Taiwanese highway traffic behavior.
We propose a prediction-based detection method for finding highway traffic anomalies using reconciled ordinary least squares (OLS) forecasts and bootstrap prediction intervals. Two fundamental features of traffic flow time series -- namely, seasonality and spatial autocorrelation -- are captured by adding Fourier terms in OLS models, spatial aggregation (as a hierarchical structure mimicking the geographical division in regions, cities, and stations), and a reconciliation step. Our approach, although simple, is able to model complex traffic datasets with reasonable accuracy. Being based on OLS, it is efficient and permits avoiding the computational burden of more complex methods. Analyses of Taiwan's consecutive holidays in 2019, 2020, and 2021 (73 days) showed strong variations in anomalies across different directions and highways. Specifically, we detected some areas and highways comprising a high number of traffic anomalies (north direction-central and southern regions-highways No. 1 and 3, south direction-southern region-highway No.3), and others with generally normal traffic (east and west direction). These results could provide important decision-support information to traffic authorities.
}


 \keywords{Traffic patterns, Traffic anomalies, Hierarchical and grouped forecasting, Reconciling forecast, Linear regression}

\maketitle

\section{Introduction}
	
	During consecutive holidays in Taiwan, the demand for the national highway is 1.5 to 2 times more than on weekdays\footnote{Chinese version reference: \url{www.freeway.gov.tw/Download_File.ashx?id=13014}}. 
	To maintain a reasonable quality of service and cope with anticipated traffic spikes during these holidays, the Taiwan freeway bureau implements various freeway controls and management strategies depending on the characteristics of the holidays.  
	Traffic authorities use different strategies for traffic management, such as planning alternative routes, flexible adjustments on signal timing, traffic directing, and enforcement by local police, ramp metering on roads connecting to freeways, and other temporary transport managements\footnote{\url{www.thb.gov.tw}}. 
	Other management strategies restrict vehicles from accessing the road network based on their occupancy or their license numbers \citep[high-occupancy or odd-even license plate policies,][]{chu2015road}. 
        However, due to the exponential growth of registered motor vehicles in Taiwan\footnote{\url{tradingeconomics.com/taiwan/car-registrations}}, controlling traffic flow is becoming more and more difficult for traffic authorities, particularly during consecutive holidays.

	Data points that deviate from a dataset normal behavior are known as anomalies or outliers. In this study we are interested in detecting anomalies which are spikes in the series, i.e.~points with extreme values compared to the neighboring points in the series. In particular, we say that the series as an anomaly at time $t$ if 
 \begin{equation}
     |x_t - \hat{x}_t|> threshold_t,
 \end{equation}
 where $x_t$ is the observed time series value, $\hat{x}_t$ is the prediction given by our model \citep{darban2022deep}, and $threshold_t$ is based on the corresponding estimated prediction interval. Specifically, we consider as anomalies the points which exceed the upper limit of the prediction intervals.
	
	Anomaly detection in highway traffic flow is a vital aspect of intelligent traffic management systems, and it has been discussed thoroughly in recent years. Indeed, anomalies on highways indicate traffic congestion which can be caused by several factors, such as accidents, rush hours, or severe weather conditions. 
	Importantly, early and accurate detection of unusual traffic events can improve smooth traffic flow on roads and avoid life and economic losses \citep{mondal2020road,tang2005traffic}. 
	
	There is a vast literature on traffic anomaly detection \citep[see, e.g.~, the survey in][]{djenouri2019survey}. 
	As an example of a recent paper on this subject, \cite{bawaneh2019anomaly} introduced a new anomaly detection algorithm that searches for significant changes in the occupancy's road traffic time series. In particular, they transformed the time series with a derivative estimation model into a symbolic representation sequence and used a modified z-score to detect the anomalies in the symbolic sequence. 
	Another example is represented by the work of \cite{huang2018traffic}, who suggested a spatiotemporal pattern network architecture to detect traffic system-level anomalies in a batch-processing fashion. They showed that their proposed technique could effectively capture time series features and discover spatial and temporal patterns for the traffic system. 
	\cite{zhang2016spatial} analyzed the large-scale traffic data by identifying its features for temporal and spatial patterns. They derived an anomaly index to quantify the network traffic in both features. 
	\cite{riveiro2017anomaly} presented a visual framework that can explore and analyze the multidimensional road traffic data. This framework also detects anomalies and shows an explanation of anomalous events. 
	Finally, a complex method for traffic anomaly detection based on deep learning was recently proposed by \cite{aboah2021vision}.  Their approach includes an anomaly detection model followed by a detection and analysis step. The model foundation is YOLOv5, and the second step includes traffic scene background estimation, road mask extraction, and adaptive thresholding. The candidate anomalies obtained by this pipeline are then passed through a decision tree to detect and analyze the final anomalies. 
	
	Prediction-based anomaly detection algorithms similar to the one presented in this paper are becoming an essential tool in several fields beyond traffic and transportation. 
	For example, \cite{hou2013detection} computed the forecast for water quality variation tendencies and detected water quality anomalies using the abnormal thresholds of prediction residuals. Another example is presented by \cite{li2019detection}, who suggested a voltage prediction-based anomaly detection algorithm for spacecraft storage batteries based on a deep belief network (DBN). 
	Finally, \cite{pang2018optimize} designed a graphic indicator of the receiver operating characteristic (ROC) curve of the prediction interval, which helps to optimize the prediction interval coverage for detecting anomalous events. 
	When this approach is applied to traffic forecast and anomaly detection, one must consider two important characteristics of traffic data such as seasonality and spatial correlations.

  In the field of time series anomaly detection, a simple and efficient method involves the application of average-based techniques, the majority of which fall into the category of prediction-based methods, as discussed in the survey by \citet{zhong2023survey}. A simple averaging technique consists in the use of Moving Averages (MAs). For example, in \citet{choffnes2010crowdsourcing}, a naive MA approach was used to detect unusual network events at the edge of the network. The Autoregressive Integrated Moving Average (ARIMA) model represents an improvement over the naive MA technique and is well-suited for accurate time series forecasting and hence for anomaly detection. For instance, \citet{zhang2005network} segmented the inference and anomaly detection stages, testing various models for identifying anomalies in time series data. Their experiments revealed the effectiveness of algorithms that integrate ARIMA with $L^1$-norm minimization when dealing with time series data containing noise and missing values. Another prediction-based approach using ARIMA for detecting anomalous points in time series is presented in \citet{qiu2019short}. The authors utilized a hybrid forecasting model called ARIMA-WNN, which combines ARIMA to model the linear component of a time series and a Wavelet Neural Network (WNN) to predict the residual nonlinear component.
 
 %

	In this paper, we propose a prediction-based method for detecting traffic anomalies by modeling and forecasting hierarchical time series, and we apply it to Taiwanese highway data, specifically focusing on consecutive holidays (e.g.,~the Chinese new year holiday), for which the government sets plans to control the traffic flow. 
	In particular, we propose to exploit an ordinary least squares (OLS) forecasting model comprising Fourier terms to reflect the seasonality, while geographical aggregation is incorporated through a hierarchical structure of data, along with a reconciliation mechanism.
	
	In addition, we propose a block cross-validation approach for daily (24 hours) forecasting to detect anomalies every 24 hours. 
	
	The paper is organized as follows. 
		Section \ref{sec:data} briefly presents the main Taiwanese highways and introduces the traffic flow time series analyzed in this work. 
		In Section \ref{sec:method}, we outline the proposed methodology for hierarchical and group time series and we detail the model specification for the application to traffic anomaly detection in Taiwanese highways. 
		The results for traffic flow forecast and consecutive holiday anomaly detection are presented in Section \ref{sec:result}. 
		Finally, we draw our conclusion in Section \ref{conclusion}.
\section{Taiwanese highways data}
	\label{sec:data}
	
\begin{figure}[!htp]
		\centering
		\subfloat[North]{\label{Taiwan-highway-map-north}\includegraphics[width=11cm,height=7cm]{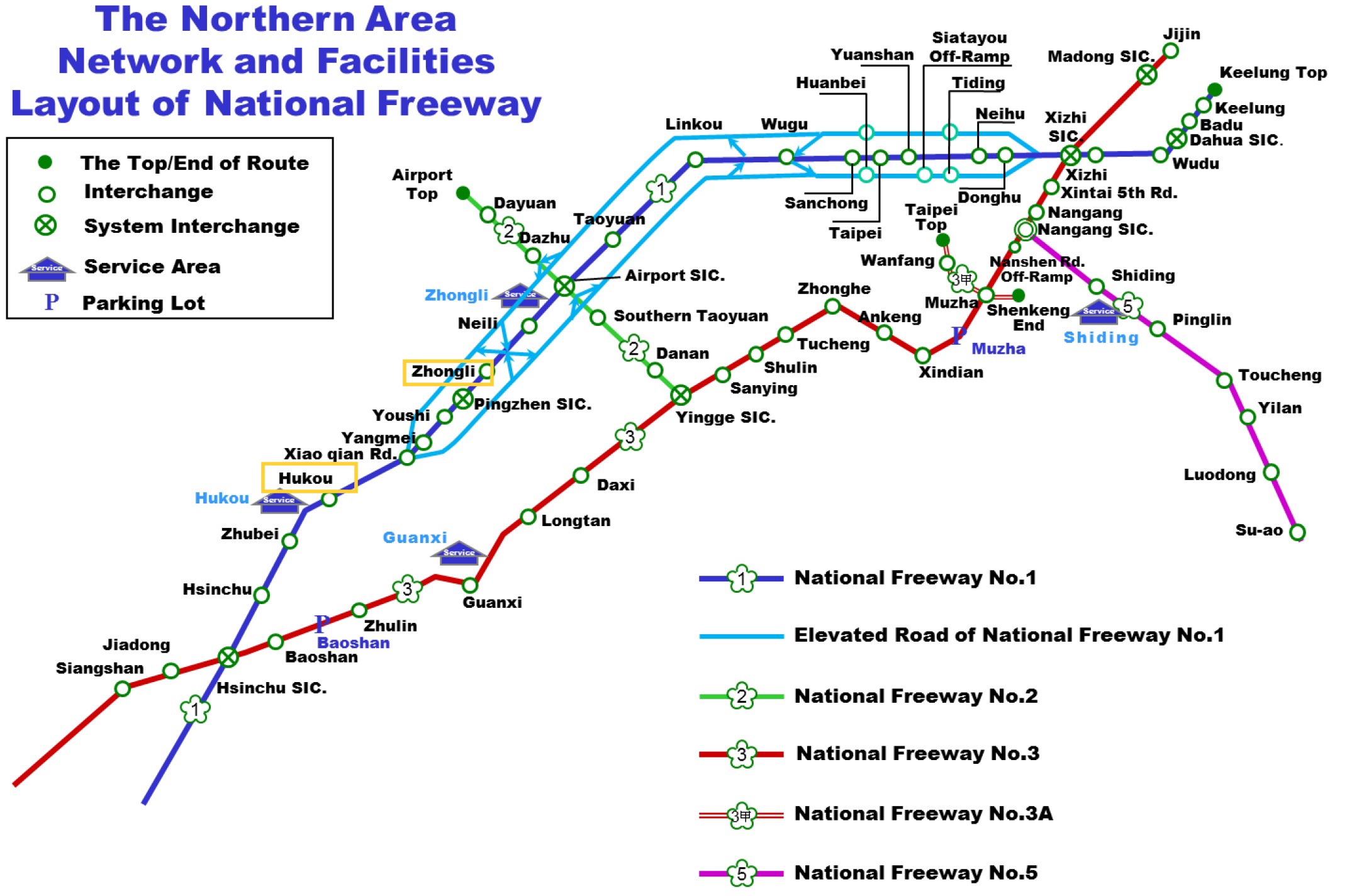}}\\
		\centering
		\subfloat[Central]{\includegraphics[width=0.48\textwidth]{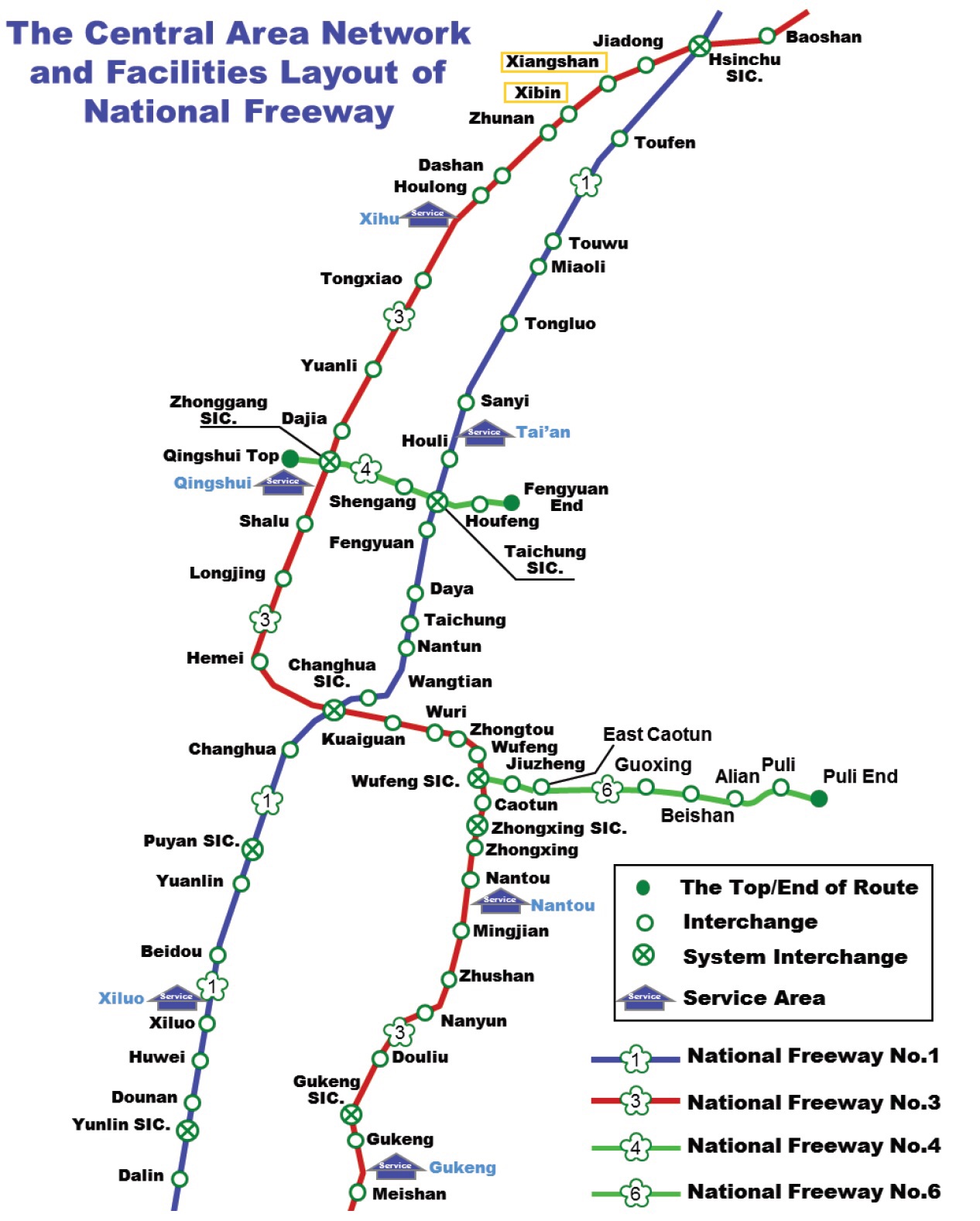}\label{Taiwan-highway-map-center}}
		\hfill
		\subfloat[South]{\includegraphics[width=0.48\textwidth]{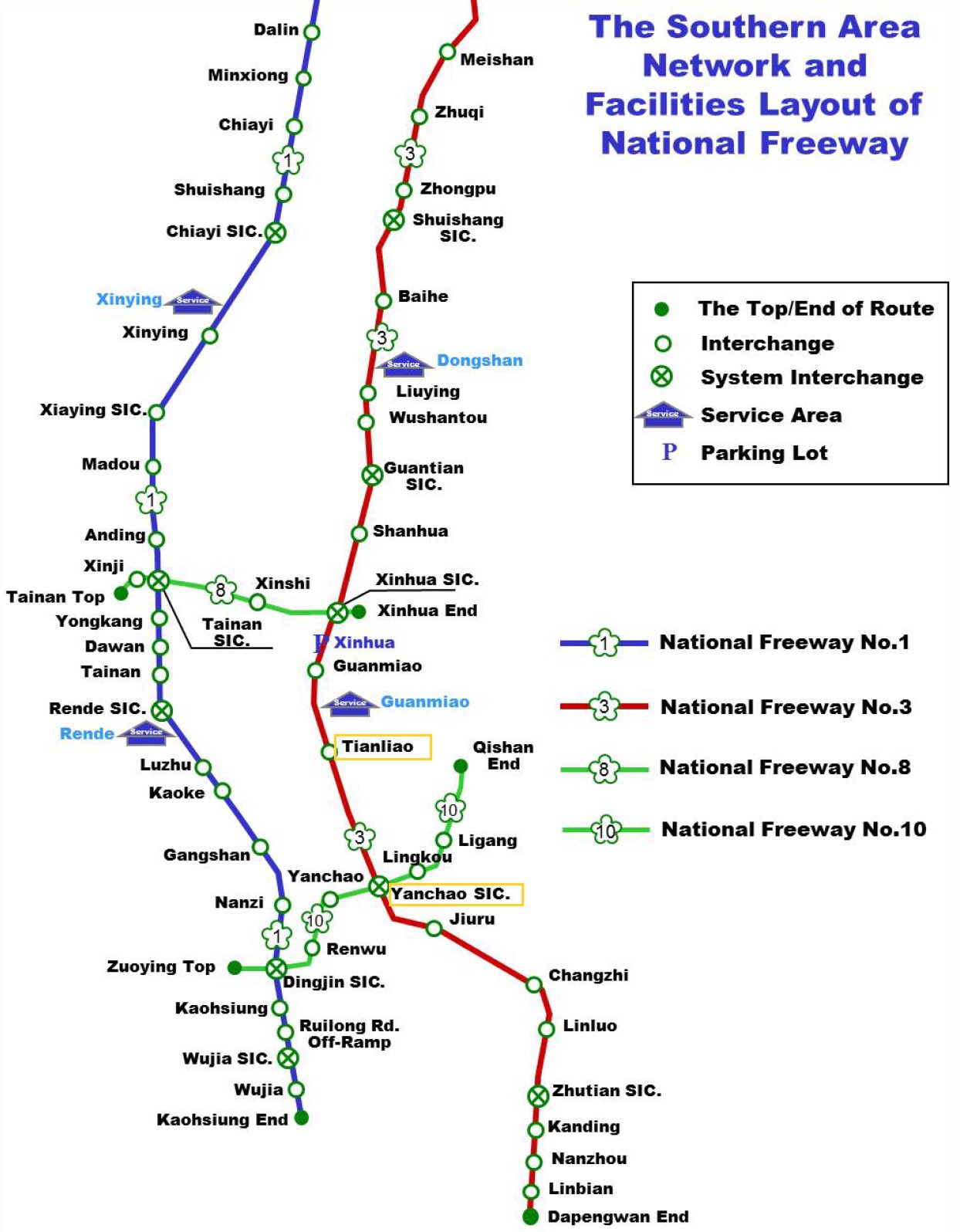}\label{Taiwan-highway-map-south}}
		\caption{
			Taiwanese highway network maps in northern, central and southern regions with indication of the sensor stations. 
			Yellow rectangles highlight the sensor stations corresponding to Figures \ref{fig:example-Taiwan-traffic-series} and \ref{fig:example-Taiwan-traffic-series-7days}.
			Figures from \protect\url{www.freeway.gov.tw/english/way_net.aspx?cnid=1110}. 
		}
		\label{Taiwan-highway-map}
	\end{figure}

	The main Taiwanese highway network involves highways 1, including its elevated part (Keelung city - Kaohsiung), 3 (Jijin, Keelung city - Dapengwan, Pingtung county), and 5 (Nangang, Taipei city - Su-ao, Yilan). 
	Figure \ref{Taiwan-highway-map} illustrates these highways on the Taiwan map in northern (Figure \ref{Taiwan-highway-map-north}), central (Figure \ref{Taiwan-highway-map-center}), and southern (Figure \ref{Taiwan-highway-map-south}) regions. Although the Taiwanese highway network includes additional highways (as displayed in Figure \ref{Taiwan-highway-map}), traffic data is available only for the three main highways mentioned above.
	There are 335 detectors located on these highways, which are part of the Electronic Tolling Collection (ETC) system obtained from the Taiwan Freeway Bureau MOTC website\footnote{\url{tisvcloud.freeway.gov.tw/history/TDCS/}}. 
	Every five minutes, the detectors in the ETC system record the number and type of vehicles that pass through it \citep{siu2020switching}. The types of vehicles considered in this study include cars, small trucks, buses/coaches, big trucks, and tractor-trailers.
	The data collected by the ETC system are publicly available and can be downloaded from the website of the Taiwan Freeway Bureau, MOTC.
	
	Our study concerns two years and four months of traffic flow in Taiwan, from 2019-01-01 to 2021-04-30, comprising 73 days of consecutive holidays (see Table \ref{tab:TCH}).
	For computational reasons, as well as to make the series smoother and reduce noise, we aggregated the data in order to obtain hourly time series (consecutive holidays correspond to $73\times24=1752$ hours). For simplicity, we exclude Highway 5 from our dataset. 
	We also exclude the ``all zero'' and ``mostly zero'' ($\geq 80\%$ of zero values) time series -- which amounts at around $1\%$ of the total number of series. 
	The final dataset includes 1590 time series at the most disaggregated level (see Subsection \ref{sec:HGTS}), 
		each with length of 20424 observations.
	
	\begin{table}[!ht]
		\caption{\label{tab:TCH}Taiwanese consecutive holidays in the time period from 2019-01-01 to 2021-04-30. The last long holiday in the dataset starts on the last day of the time period considered in the study, hence it corresponds to only one day.}
		\centering
		\begin{tabular}[t]{lr}
			\toprule
			Year & Consecutive holidays\\
			\midrule
			2019 & \\
			& 02-02 to 02-10 (9 days) \\
			& 02-28 to 03-03 (4 days) \\
			& 04-04 to 04-07 (4 days) \\
			& 06-07 to 06-09 (3 days) \\
			& 09-13 to 09-15 (3 days) \\
			& 10-10 to 10-13 (4 days) \\
			\hline
			2020 & \\
			& 01-23 to 01-29 (7 days)\\
			& 02-28 to 03-01 (3 days) \\
			& 04-02 to 04-05 (4 days) \\
			& 05-01 to 05-03 (3 days) \\
			& 06-25 to 06-28 (4 days) \\
			& 10-01 to 10-04 (4 days) \\
			& 10-09 to 10-11 (3 days) \\
			\hline
			2021 & \\
			& 01-01 to 01-03 (3 days)\\
			& 02-10 to 02-16 (7 days) \\
			& 02-27 to 03-01 (3 days) \\
			& 04-02 to 04-05 (4 days) \\
			& 04-30 to 04-30 (1 day) \\
			\hline
			Total  & 73 days\\
			\bottomrule
		\end{tabular}
	\end{table}
	
\begin{figure}[!htp]
		\centering
		\subfloat[Northern region, Zhongli runway - Hukou, south direction, car]{\label{N2971S31}\includegraphics[width=0.92\textwidth]{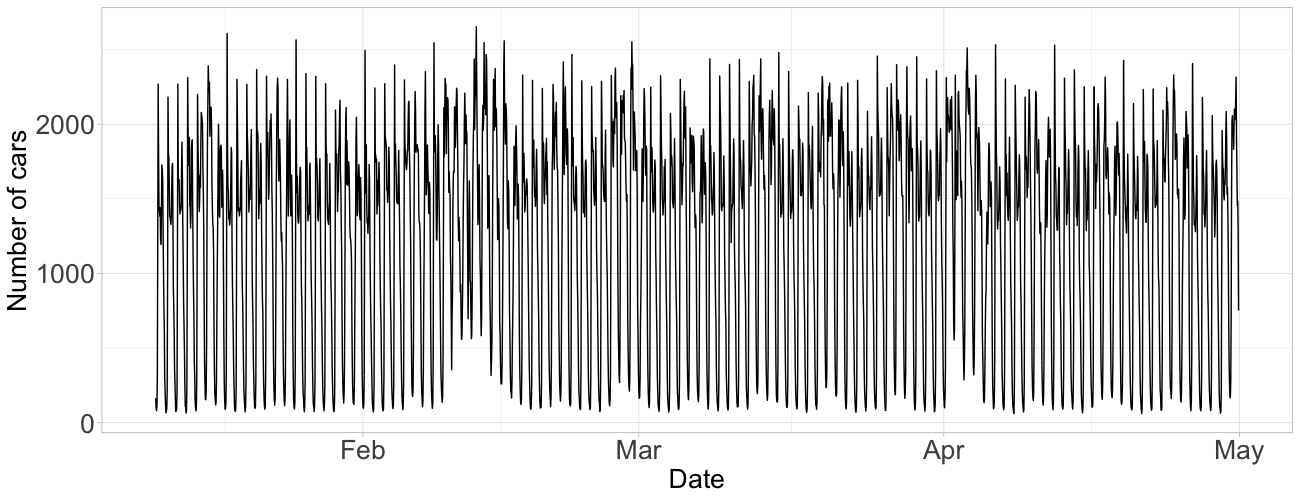}}\\
		\centering
		\subfloat[Central region, Xiangshan - Xibin, south direction, small truck]{\label{C2443S32}\includegraphics[width=0.92\textwidth]{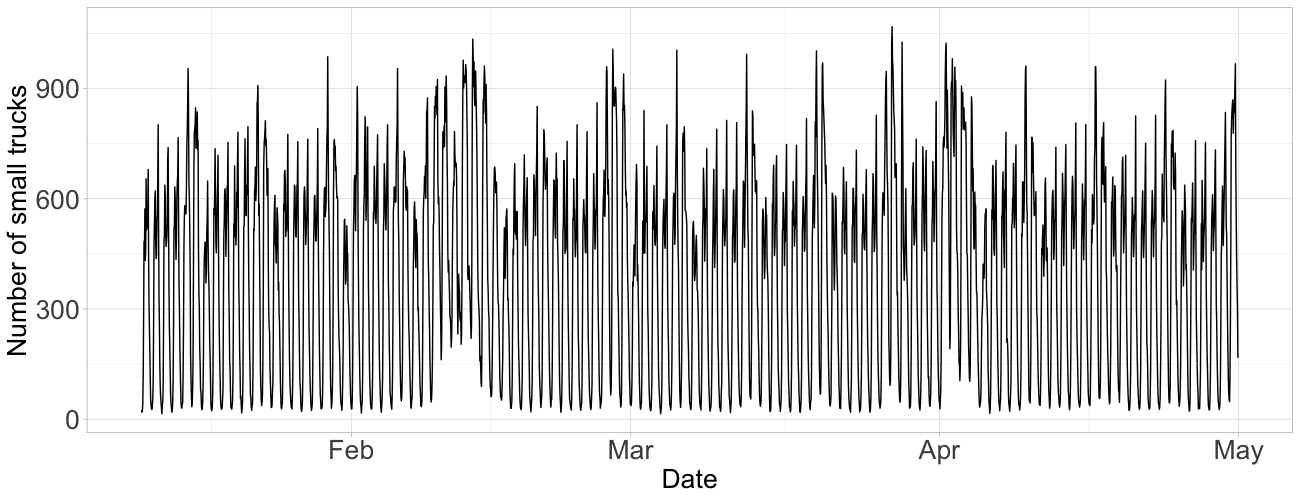}}\\
		\centering
		\subfloat[Southern region, Yanchao SIC - Tianliao, north direction, big truck]{\label{S2693N42}\includegraphics[width=0.92\textwidth]{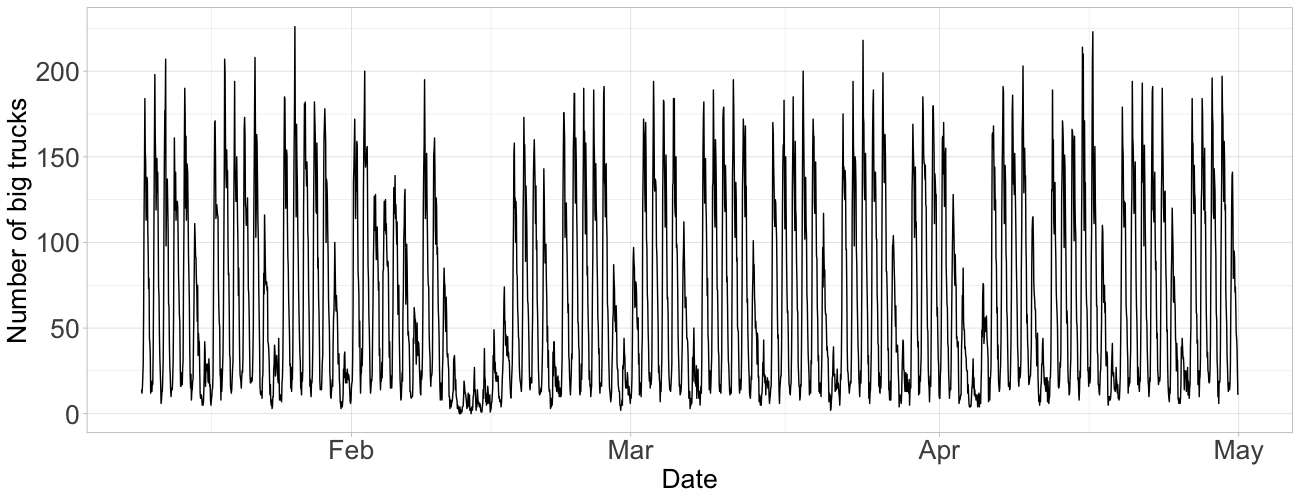}}
		\caption{
			Examples of Taiwanese highway hourly time series (from 2021-01-10 23:00:00 to 2021-04-30 22:00:00) in three regions for different stations, traffic directions, and vehicle types. 
		}
		\label{fig:example-Taiwan-traffic-series}
	\end{figure}

\begin{figure}[!htp]
		\centering
		\subfloat[Northern region, Zhongli runway - Hukou, south direction, car]{\label{N2971S31-7days}\includegraphics[width=0.75\textwidth]{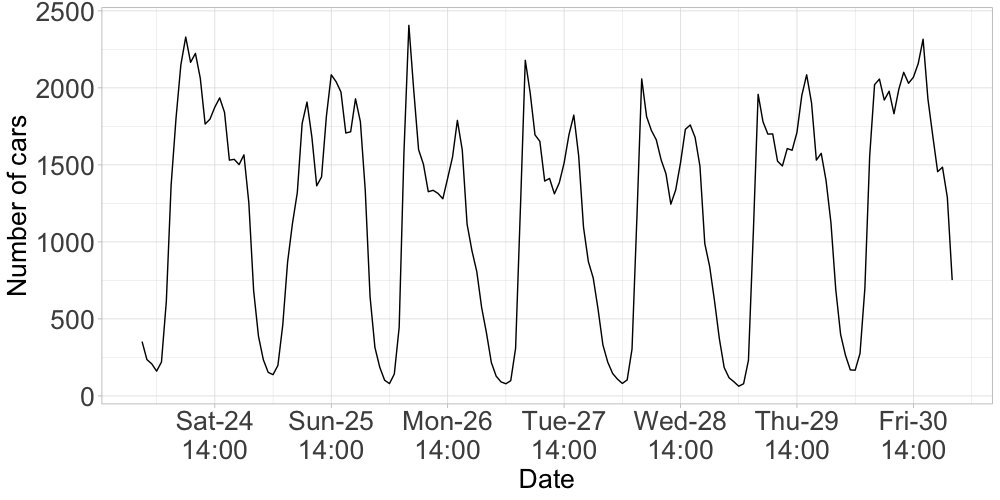}}\\
		\centering
		\subfloat[Central region, Xiangshan - Xibin, south direction, small truck]{\label{C2443S32-7days}\includegraphics[width=0.75\textwidth]{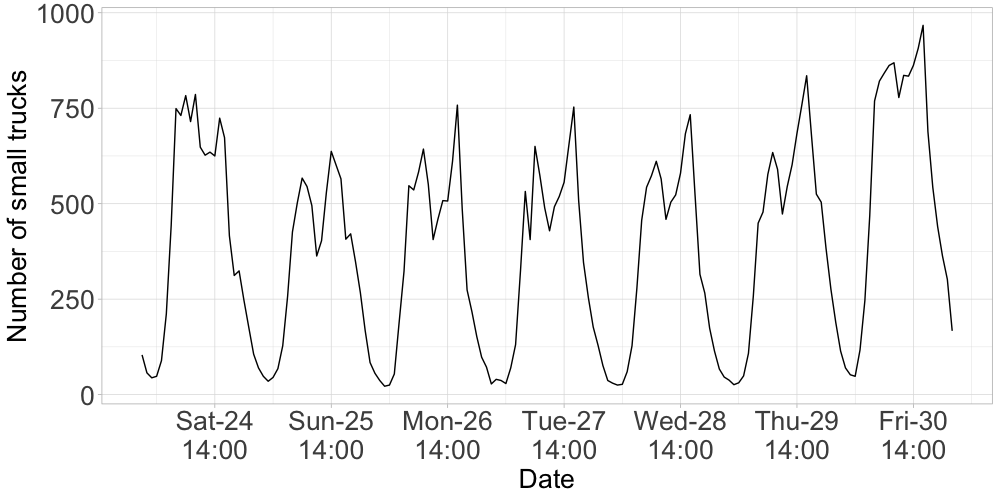}}\\
		\centering
		\subfloat[Southern region, Yanchao SIC - Tianliao, north direction, big truck]{\label{S2693N42-7days}\includegraphics[width=0.75\textwidth]{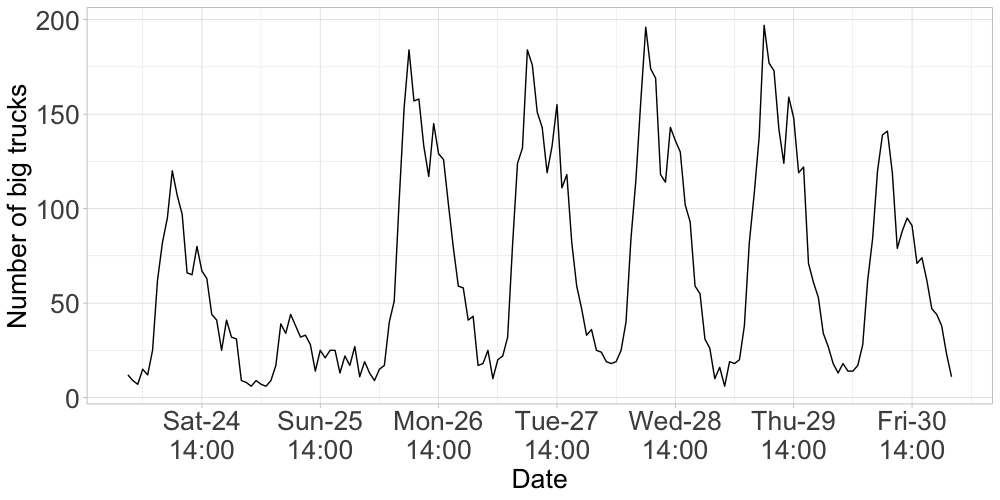}}
		\caption{
			Examples of Taiwanese highway hourly time series, zoomed in view on the last seven days of the time series shown in Figure \ref{fig:example-Taiwan-traffic-series} (from 2021-04-23 23:00:00 to 2021-04-30 22:00:00).
		}
		\label{fig:example-Taiwan-traffic-series-7days}
	\end{figure}

	Figure \ref{fig:example-Taiwan-traffic-series} displays three examples of Taiwan highway hourly traffic time series corresponding to different vehicle types in different regions, stations, and traffic directions (for four months, from 2021-01-10 23:00:00 to 2021-04-30 22:00:00). 
	The corresponding sensor stations are highlighted with yellow rectangles in \mbox{Figure \ref{Taiwan-highway-map}}. 
	Figure \ref{fig:example-Taiwan-traffic-series-7days} provides a zoomed-in view on the last seven days for the same three time series (hence, from 2021-04-25 23:00:00 to 2021-04-30 22:00:00). 
	In both figures, we can clearly see daily patterns in the Taiwan traffic flow (peak flow in the daytime and much lower flow at night). 
	More in detail, Figure \ref{fig:example-Taiwan-traffic-series-7days} shows that in northern and central regions, the first five days exhibit highly similar daily patterns, with a slightly increased flow on Friday and the weekend.
    Conversely, in the southern region, there is a decrease in big track flow on Friday and Saturday and a significant drop on Sunday.
	In addition, we can observe weekly patterns in Taiwan traffic flow (Figure \ref{fig:example-Taiwan-traffic-series}). In particular, the weekly patterns in the first four weeks are similar within each time series, while we observe peaks during the fifth week (mid-February) in the series of Figures \ref{N2971S31} and \ref{C2443S32}, indicating a higher number of cars and small trucks in these highways, respectively. Interestingly, these peaks correspond to a valley in the time series of Figure \ref{S2693N42}, which represents the flow of large trucks heading north from station Yanchao SIC to Tianliao in the southern region. 
	This could be due to the Chinese New Year holidays and the imposition of a restriction on large trucks in that particular area of the southern region.

	\section{Methods}\label{sec:method}
	
	In order to perform anomaly detection in Taiwan traffic time series, we propose to adapt and employ the ordinary least squares (OLS) forecasting model under the hierarchy and grouped framework developed by \cite{ashouri2021fast}. 
	In this paper, we refine the OLS model of \cite{ashouri2021fast} by incorporating Fourier terms in order to capture the seasonal patterns on a daily and weekly basis in our traffic dataset, while we employ the hierarchy and grouping among traffic time series to embed their spatial correlation and geographical aggregation.
	In contrast to \cite{ashouri2021fast}, where the prediction intervals were derived based on the normality assumption, here we construct bootstrap prediction intervals that only require uncorrelated forecast errors. 
	Finally, we employ these intervals for performing prediction-based anomaly detection \citep{pang2018optimize} on the Taiwan traffic time series, i.e.~, we label a point as an anomaly if the actual traffic flow value exceeds the prediction interval's upper bound. 
	The rest of this section details our modeling and anomaly detection framework. 
	
	\subsection{Hierarchical and grouped time series} \label{sec:HGTS}
	
\begin{figure}[!htp]
		\centering
		\includegraphics[width=0.4\textwidth]{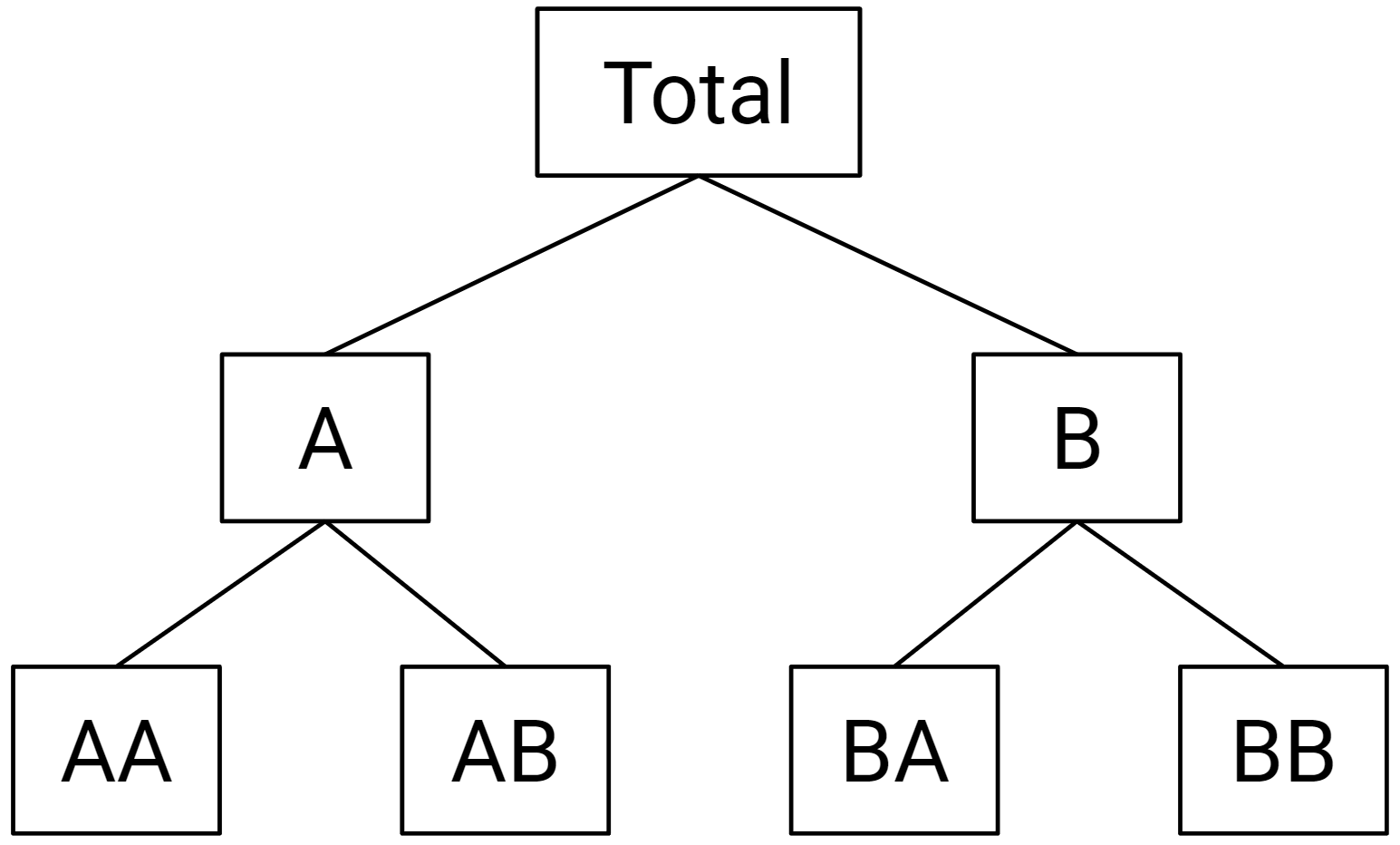}
		\caption{Example of two-level hierarchy structure.}
		\label{fig:hierarcy-example}
	\end{figure}

	In several real-data applications, a large collection of related time series is structured in a hierarchical and/or grouped framework. For instance, traffic data can be disaggregated based on a geographical hierarchy of regions, cities, and stations. 
	Figure \ref{fig:hierarcy-example} shows an example of a two-level hierarchy structure. The upper level of the hierarchy (level 0), i.e., the $Total$ series, represents the most aggregated time series. At level 1, the $Total$ series is disaggregated into series $A$ and $B$. Finally, the series $A$ is disaggregated into series $AA$ and $AB$, while $B$ into series $BA$ and $BB$, at level 2. The most disaggregated level of the hierarchy structure -- which is level 2 in the example -- is called the bottom-level series. 
	The grouped structure is similar to the hierarchical structure, except that in this case the structure does not disaggregate in a unique hierarchical way \citep[for more details, see][]{fpp3}. 
	
	Following the notation in \cite{fpp3}, we indicate with $y_t$ the $Total$ series for $t = 1, 2, ..., T$, where $T$ represents the number of time points of the series. 
	For the other series in the hierarchy structure, $y_{jt}$ represents the $t^{\text{th}}$ observation of the series corresponding to node $j$. 
	For example, in Figure \ref{fig:hierarcy-example}, $y_{At}$  and $y_{BBt}$ indicate the $t^{\text{th}}$ observation in node $A$ at level 1 and $BB$ at level 2, respectively. 
	In addition, we indicate with $\bf S$ the {\it summing matrix}, which represents how bottom-level series are aggregated. 
	Let $n$ be the total number of series in the hierarchy, $m$ the number of bottom-level series, ${{\bf y}_t}$ the $n$-vector of all the series in the hierarchy, and ${{\bf b}_t}$ the $m$-vector of all the bottom-level series; the summing matrix $\bf S$ is the $n\times m$ matrix such that ${{\bf y}_t} = {\bf S} {{\bf b}_t}$. 
	For example, the following equation shows the $\bf S$ matrix for the structure in Figure \ref{fig:hierarcy-example} (in this case, $n = 7$ and $m = 4$):
	
	\begin{equation*}
		\begin{pmatrix}
			y_{t}\\y_{A,t}\\y_{B,t}\\y_{AA,t}\\y_{AB,t}\\y_{BA,t}\\y_{BB,t}
		\end{pmatrix} =
		\begin{pmatrix}
			1&1&1&1\\1&1&0&0\\0&0&1&1\\1&0&0&0\\0&1&0&0\\0&0&1&0\\0&0&0&1\\
		\end{pmatrix}
		\begin{pmatrix}
			y_{AA,t}\\y_{AB,t}\\y_{BA,t}\\y_{BB,t}\\
		\end{pmatrix}.
	\end{equation*}
	
	Importantly, the forecasts in all the hierarchy levels should add up properly. 
	Indeed, if the aggregation constraints are ignored, then the resulting forecasts would not be consistently aggregated and the relationships between the series would be missed \citep{fpp3}. 
	The methods traditionally used to forecast hierarchical and grouped time series are called single-level approaches and include bottom-up \citep{kahn1998revisiting}, top-down \citep{gross1990disaggregation}, and middle-out \citep{hyndman2011optimal} approaches. All these methods select one level of aggregation and generate the forecasts for all series at that level; they then aggregate the forecasts for higher level series and disaggregate them for lower level series. 
	In this paper, we employ a more recent approach called optimal combination which includes two steps. First, the forecasts of all the series in all the levels of aggregation are computed ({\it base forecasts}); second, the base forecasts are reconciled ({\it reconciled forecasts}). 
	Let ${\bf {\hat y}}_{t+h}$ indicate the $h$-step-ahead base forecasts; then, the reconciled forecast ${\bf {\tilde y}}_{t+h}$ can be computed by the following equation
	\begin{equation}\label{eq:reconciledF}
		{{\bf {\tilde y}}_{t+h}} = {\bf S} ({\bf S}'{{\bf W}_h}^{-1}{\bf S})^{-1} {\bf S}' {{\bf W}_h}^{-1} {{\bf {\hat y}}_{t+h}} = {\bf S}{\bf G} {{\bf {\hat y}}_{t+h}},
	\end{equation}
	where $\bf S$ is the summing matrix, ${{\bf W}_h} = Var({{\bf y}_{t+h}} - {{\bf \hat{y}}_h})$, and ${\bf S}{\bf G}$ is the {\it reconciliation matrix}. 
	Since estimating  ${{\bf W}_h}$ from data is generally challenging, some approximations have been proposed in the literature. In this paper, we apply the {\it shrinkage estimator} \citep{wickramasuriya2019optimal}, which shrinks the full covariance matrix (of the errors on one-step predictions made on the training data) toward a diagonal matrix.
	
	In our Taiwan traffic dataset, the hierarchy and grouped structure are defined by geographic divisions, highway number, direction, and type of vehicle. In particular, the first geographic division is represented by the three network regions, i.e.,~``north'', ``center'', and ``south''. The second level in the hierarchical structure divides the dataset into 319 stations 
	(see Figure \ref{Taiwan-highway-map}). 
	In addition, the data has a grouped structure with the three following attributes:
	\begin{enumerate}
		\item Highway number: no.1, elevated no.1, and no.3;
		\item Highway direction: north, south, east, and west;
		\item Vehicle type: car, small truck, bus/coach, big truck, and tractor-trailer.
	\end{enumerate}
	This structure results in a total of 1998 series, when considering all levels.
	Table \ref{tab:trafficdivision} displays all the aggregation levels and the combinations applied in the reconciliation step (see Subsection \ref{sec:OLS}), with the number of series at each hierarchy level. 
	Note that, for simplicity, we only include two-way combinations in the reconciliation step. Including also three-way combinations could slightly improve the results. 
	
	\begin{table}[ht]
		\caption{\label{tab:trafficdivision}Number of highway traffic series at each aggregation level.}
		\centering
		\begin{tabular}[t]{lr}
			\toprule
			Aggregation level & Series\\
			\midrule
			Taiwan & 1\\
			Region & 3\\
			Station & 319\\
			Highway & 3\\
			Direction & 4\\
			Vehicle type & 5\\
			Region x Direction & 8\\
			Region x Highway & 7\\
			Region x Vehicle type & 15\\
			Highway x Direction & 8\\
			Highway x Vehicle type & 15\\
			Direction x Vehicle type & 20\\
			Bottom-level series & 1590\\
			\hline
			Total & 1998\\
			\bottomrule
		\end{tabular}
	\end{table}
	
	Other fundamental features of our traffic data that must be included in the model are the complex seasonal patterns (daily and weekly seasonality), the autocorrelation, and the spatial correlations \citep[traffic volumes at nearby locations are highly correlated;][]{li2019short}. The OLS approach we propose includes specific terms to handle autocorrelation and multiple seasonalities (see Subsection \ref{sec:OLS}), while spatial correlation is already embedded in the hierarchical structure of the series and is taken care of in the reconciliation step \citep{wickramasuriya2019optimal}. 
	For previous research on traffic forecast, we refer to \cite{li2019short}, who combined the gradient boosting procedure with hierarchical reconciliation for short-term traffic flow forecasting. In contrast to the approach proposed in \cite{li2019short}, which involves a complex method that may encounter computation difficulties, our approach employs a straightforward and interpretable linear model that performs effectively on intricate highway problems.

	\subsection{Ordinary least squares forecast reconciliation}\label{sec:OLS}
	
	To calculate the base forecasts (${\bf\hat{y}}_{t+h}$), we employ OLS models with relevant predictors for forecasting hourly traffic time series. 
	While being very simple and computationally efficient, OLS models can provide very good approximations of more complex models' forecasts \citep{ashouri2018assessing,ashouri2019tree}. Moreover, the computational benefits of the OLS approach allow us to use bootstrap to compute the prediction intervals, avoiding specific distribution assumptions. 
	Our predictor set includes linear trend, seasonality, and lag variables. 
	In particular, we assume daily and weekly seasonality, hence we include Fourier terms \citep{fpp3} for each of these two seasonal periods (periods of $24$ and $7 \times 24 = 168$ hours for daily and weekly seasonality, respectively). 
	To capture time series autocorrelation, we include the first and the $24^\text{th}$ lags among our predictors (see Appendix \ref{app:ACFPACF} for details on the selection of terms for the OLS model). 
	Hence, the proposed OLS model is
	
	\begin{equation}\label{eq:linearmodel}
		\begin{aligned}
			y_t = \alpha_0 + \alpha_1 t + \sum_{k_1=1}^{l_1}\beta_{k_1} \sin\left(\frac{2\pi k_1 t}{24}\right) + \sum_{k_1=1}^{l_1}\beta'_{k_1} \cos\left(\frac{2\pi k_1 t}{24}\right) \\
			+ \sum_{k_2=1}^{l_2}\gamma_{k_2} \sin\left(\frac{2\pi k_2 t}{7\times24}\right) + \sum_{k_2=1}^{l_2}\gamma'_{k_2} \cos\left(\frac{2\pi k_2 t}{7\times24}\right) \\
			+ \eta_1 y_{t-1} + \eta_{24} y_{t-24} + \varepsilon_t,
		\end{aligned}
	\end{equation}
	where \(y_{t-k}\) is the \(k^\text{th}\) lagged value for \(y_t\), $l_1=3$, $l_2=2$, and $\varepsilon_t$ is the error term. 
	For computing base forecasts, one OLS model with the same set of predictors in Equation \ref{eq:linearmodel} is fitted to every single series in the hierarchical and grouped structure. 
	The forecast is then reconciled by using Equation \ref{eq:reconciledF}. 
	Algorithm \ref{alg:algorithmBF} shows the procedure we propose for calculating base and reconciled forecasts. 
	
	\begin{algorithm}[h]
		\begin{algorithmic}[1]
			\caption{Computing base and reconciled forecasts using OLS models}\label{alg:algorithmBF}
			\STATE Forecast horizon : $h$
			\FOR{$i \in \{1,\dots,N\}$}
			\STATE Time series : ${{\bf y}_i} = (y_{1,i}, y_{2,i}, \dots, y_{T,i})$
			\STATE $P$ predictors: ${{\bf X}_i} =\{(x_{1,i,1}, x_{2,i,1}, \dots, x_{T,i,1}), \dots, (x_{1,i,p}, x_{2,i,p}, \dots, x_{T,i,p})\}$
			\STATE Model: $fit = lm(y_{t,i} \sim x_{t,i,1} + x_{t,i,2} + \dots + x_{t,i,p}, ~data = ({\bf y}_i, {\bf X}_i))$
			\STATE $P$ new predictors: ${{\bf X}_i}^{\dag} =\{(x_{T+1,i,1}, x_{T+2,i,1}, \dots, x_{T+h,i,1}), \dots, (x_{T+1,i,p}, x_{T+2,i,p}, \dots, x_{T+h,i,p})\}$
			\FOR{$j \in \{1,\dots,h\}$}
			\STATE Prediction: $\hat{y}_{t+j,i} = predict(fit, ~newdata = {{\bf X}_{t+j,i}}^{\dag})$
			\ENDFOR
			\ENDFOR
			\STATE Reconcile forecasts: ${\bf\tilde{y}} = {\bf SG\hat{y}}$
		\end{algorithmic}
	\end{algorithm}

 	\begin{figure}[!htp]
		\centering
		\includegraphics[width=0.58\textwidth]{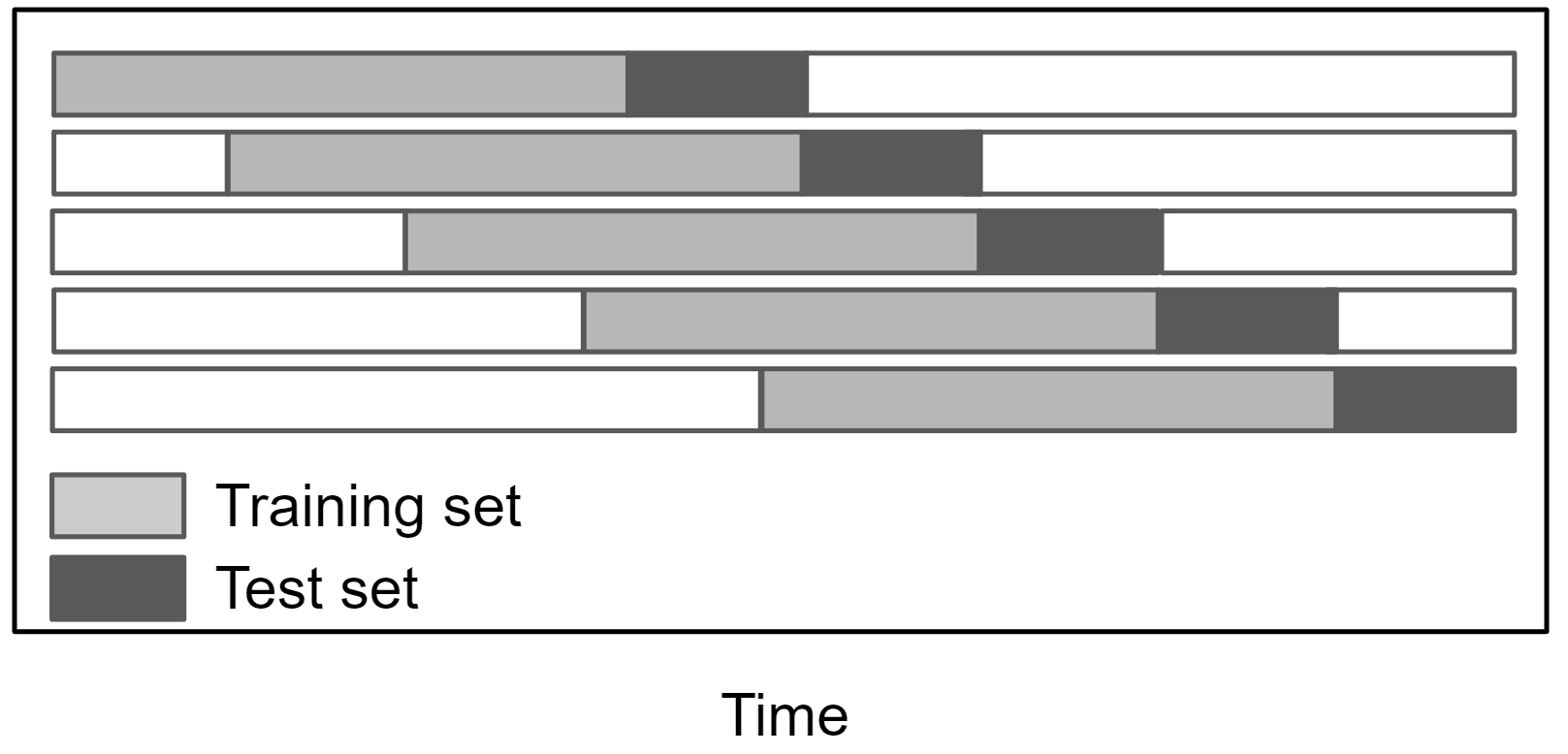}
		\caption{Blocked cross-validation approach for computing forecasts in Taiwan's traffic flow time series.}
		\label{fig:BCV}
	\end{figure}
 
	To forecast traffic flow and evaluate the performance of our approach on Taiwan traffic data (see Subsection \ref{sec:forecastingResult}), we employ a blocked cross-validation approach. 
		This approach uses a sliding window of a fixed size for selecting training and test data within a time series, respecting the order of the values in the time series. 
		In particular, the first cross-validation split considers the starting $T_{train}$ time points of the series as a training set and the next $T_{test}$ time points as a test set. Then, the test set time points are added to the training set, while an equal amount of points are removed from the beginning of the training set; the new test set is constituted by the next $T_{test}$ time points. The procedure is repeated until the test set reaches the end of the time series. 
		Figure \ref{fig:BCV} displays this procedure for creating training (light gray) and test (dark gray) sets. 
		Importantly, this approach allows us to compute the forecasts for the whole series (except the first $T_{train}$ time points).

	\subsection{Bootstrapped prediction intervals}\label{sec:BPI}
	
	Bootstrapped prediction intervals are useful when the normality assumption for the error term is unreasonable, and we can only assume uncorrelated forecast errors with constant variance. 
    In this paper, we forecast the count of vehicles per hour, hence it is advisable to drop the assumption of normality and apply the bootstrap approach\footnote{\url{ibarraespinosa.github.io/VEINBOOK/traffic.html}}.
    For this reason, we propose the Algorithm \ref{alg:algorithmBPI} to compute the bootstrapped reconciled prediction interval for our traffic data. 
	
	The proposed approach starts by estimating the OLS coefficients and computing the residuals $e_j$ for $j = 1, 2, ..., T$ for each series. 
	Then, these residuals are modified as $\{s_1-\bar{s}, s_2-\bar{s}, ..., s_T-\bar{s}\}$ where $\bar{s}$ is $s_j$'s average for $j = 1, 2, ..., T$ and
	\begin{equation}\label{eq:modifiedFE}
		s_j = \frac{e_j}{\sqrt{1-h_j}}, 
	\end{equation}
	with $h_j$ the leverage of the $j^\text{th}$ observation. 
	Note that these {\it modified residuals} are employed in order to adjust the residuals' variance differences and make their variance constant \citep{davison1997bootstrap}. 
	In the next step, a bootstrap sample is randomly drawn from the modified residuals for each forecast point and it is added to the forecast (sample path). 
	This step is repeated $K$ times, where $K$ represents the number of bootstrap samples. 
	Finally, the reconciled prediction interval is computed by reconciling the sample paths and then calculating the required percentiles \citep[see steps 17 and 18 in Algorithm \ref{alg:algorithmBPI};][]{fpp3}.

	\begin{algorithm}[ht]
		\begin{algorithmic}[1]
			\caption{Computing bootstrapped reconciled prediction intervals}\label{alg:algorithmBPI}
			\STATE Number of series: $N$
			\STATE Length of each series: $T$
			\STATE Forecast horizon : $h$
			\STATE Number of bootstrap samples: $K$
			\FOR {$i \in \{1,\dots,N\}$}
			\STATE Base forecast model: ${\bf y_i} = {\bf X_i B} + {\bf \epsilon_i}$
			\STATE  Estimate the coefficients: ${\bf {\hat B}} = ({\bf X_{i}}'{\bf X_{i}})^{-1}{\bf X_{i}}'{\bf {y}_{i}}$
			\STATE Modified residuals: ${\bf s} = \{s_1-\bar s, s_2-\bar s, ..., s_T-\bar s\}$ from model's residuals ${\bf e} = \{e_1, e_2, ..., e_T\}$
			\FOR {$j \in \{1,\dots,h\}$}
			\STATE Compute forecasts: ${\bf \hat{y}_{ij}} = {\bf X_{ij}}^{\dag}{\bf \hat B}$ 
			\FOR{$k \in \{1,\dots,K\}$}
			\STATE Draw one sample from ${\bf s}$: ${\bf s_k^*}$
			\STATE Generate future values for $i^\text{th}$ time series ($k^\text{th}$ sample path): ${\bf \hat{y}_{ij}^{[k]}} = {\bf X_{ij}}^{\dag}{\bf {\hat B}^*} + {\bf s_k^*}$ 
			\ENDFOR
			\ENDFOR
			\ENDFOR
			\STATE Reconcile the sample paths: (${\bf SG\hat{y}^{[1]}}, {\bf SG\hat{y}^{[2]}}, ..., {\bf SG \hat{y}^{[K]}}$)
			\STATE Compute pointwise $\frac{\alpha}{2}$ and $1-\frac{\alpha}{2}$ percentiles from reconciled sample paths: $p^{\frac{\alpha}{2}}_t$ and  $p^{1-\frac{\alpha}{2}}_t$
			\STATE $(1-\alpha)\%$ bootstrapped prediction interval: $(p^{\frac{\alpha}{2}}_t, p^{1-\frac{\alpha}{2}}_t)$
		\end{algorithmic}
	\end{algorithm}

	\section{Results}\label{sec:result}
	
	We present the results of two analyses of Taiwan traffic data. 
	First, we model and forecast four months of Taiwanese highway time series (from 2021-01-01 to 2021-04-30) using the OLS forecasting model under the hierarchy and grouped structure proposed in Subections \ref{sec:HGTS} and \ref{sec:OLS}, and we compare the results to the ones obtained with Autoregressive Integrated Moving Average (ARIMA) model. This analysis allows us to evaluate the performance of the proposed OLS model prior to its application in detecting traffic anomalies in consecutive holidays.
	Second, we focus on the 73 days of consecutive holidays of Table \ref{tab:TCH} and we perform anomaly detection using the method presented in Subsections \ref{sec:OLS} and \ref{sec:BPI}. 
	
	\subsection{Forecasting highway traffic flow}\label{sec:forecastingResult}
	
	We consider Taiwan's hourly traffic data from 2021-01-01 to 2021-04-30, and we model them using the proposed OLS approach in order to evaluate the forecast performance of this simple and computationally efficient model and to compare it to the forecasting results obtained by the ARIMA model. We considered another suitable method, the Exponential smoothing state space model with Box-Cox transformation \citep[TBATS;][]{fpp3}. However, we dropped it because it is too computationally expensive to be employed on our traffic dataset. 
        The ARIMA model is fitted using the function \texttt{auto.arima} from the R package \texttt{forecast} \citep{hyndman2008automatic}, with default options, while for the OLS model, we employ the \texttt{lm} function from the \texttt{stats} R package. Note that \texttt{auto.arima} function automatically selects the best ARIMA model for each time series according to AICc, while OLS model is fixed (see Subsection \ref{sec:OLS}).
        For both approaches, we include the same Fourier terms for the two seasonal periods (daily and weekly seasonalities, see Subsection \ref{sec:OLS}), as well as the same hierarchical and group structure. 
        Following the blocked cross-validation approach (Subsection \ref{sec:OLS}), we partition the data into 106 training and test set pairs, comprising 336 hours ($T_{train}$; two weeks) and 24 hours ($T_{test}$, one day), respectively. Prediction intervals are generated based on 2000 bootstrap samples. 
	
	\begin{figure}[!htp]
		\centering
		\includegraphics[width=1\textwidth]{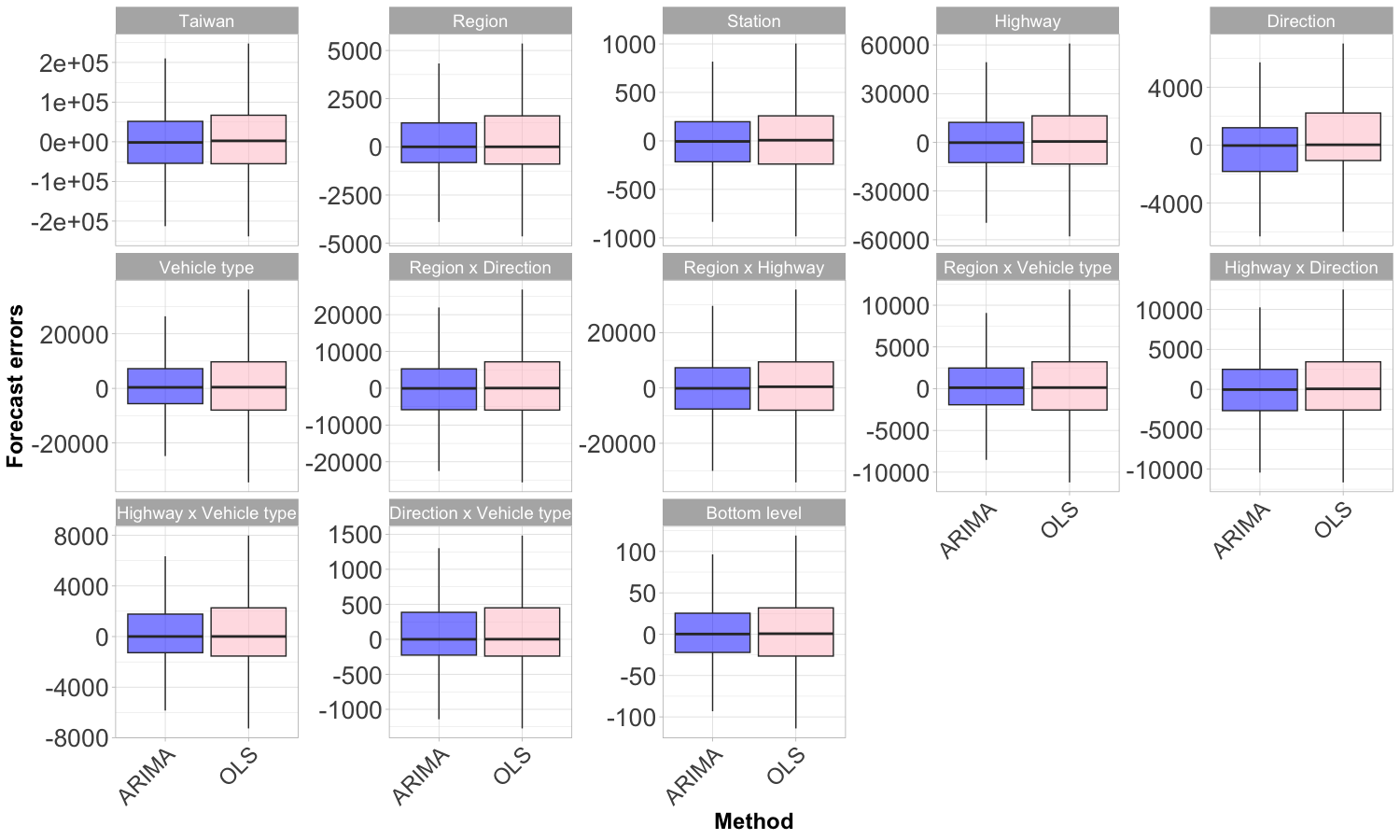}
		\caption{Boxplots of forecast errors from ARIMA and OLS methods at each aggregation level.}
		\label{fig:errorboxplot}
	\end{figure}
	
	\begin{table}[!tb]
		\caption{\label{tab:compareRMSE} Average and standard error (in parenthesis) of RMSE on test sets for ARIMA and OLS. Cells corresponding to the best average RMSE are highlighted in gray.}
		\centering
		\begin{tabular}[t]{lrr}
			\toprule
			Aggregation level &  ARIMA & OLS\\
			\midrule
			Taiwan & \cellcolor{gray!20}107038 (2120)& 115232 (2279)\\
			Region & 22432 (72)& \cellcolor{gray!20} 21533 (69)\\
			Station & \cellcolor{gray!20} 534 (0.59) & 546 (0.60)\\
			Highway & \cellcolor{gray!20} 45211 (517)& 46911 (536)\\
			Direction & \cellcolor{gray!20} 41272 (409)& 43070 (426)\\
			Vehicle type & 51083 (453)& \cellcolor{gray!20} 48236 (427)\\
			Region x Direction & \cellcolor{gray!20} 19975 (140)& 20175 (141)\\
			Region x Highway & \cellcolor{gray!20} 20122 (151)& 20221 (151)\\
			Region x Vehicle type & 18918 (67)& \cellcolor{gray!20} 17699 (91)\\
			Highway x Direction & \cellcolor{gray!20} 21405 (150)& 21722 (152)\\
			Highway x Vehicle type & 20851 (107)& \cellcolor{gray!20} 19414 (99)\\
			Direction x Vehicle type & 19104 (85)& \cellcolor{gray!20} 17854 (79)\\
			Bottom-level series & 221 (0.11)& \cellcolor{gray!20} 214 (0.11)\\
			\bottomrule
		\end{tabular}
	\end{table}

	Figure \ref{fig:errorboxplot} shows the comparison of the forecast error distributions for ARIMA and OLS approaches, for all the aggregation levels, while Table \ref{tab:compareRMSE} reports the average Root Mean Square Errors (RMSEs) and its standard error across all the series in each aggregation level. 
	Note that the errors are larger in magnitude for higher count series (i.e.,~higher aggregation levels). 
	We observe that the two methods behave similarly, without substantial bias in all aggregation levels. Interestingly, the OLS model behaves better than the ARIMA model, on average, in about half of the aggregation level, despite being much simpler and despite the fact that the best ARIMA model is automatically selected for each time series.

 	\begin{table}[!tb]
		\caption{\label{tab:computationtime} Computation time (seconds) for ARIMA and OLS with reconciliation for one cross-validation split.}
		\centering
		\begin{tabular}[t]{lrr}
			\toprule
			& ARIMA & OLS \\
			\midrule
			Base forecast & 3672 & 179\\
			Reconciling forecasts & 12 & 12\\
			Reconciling sample paths & 894 & 894\\
			\bottomrule
		\end{tabular}
	\end{table}
 
	\begin{figure}[!pht]
		\centering
		\subfloat[Taiwan]{\label{Totalolsarima}\includegraphics[width=0.95\textwidth, height=0.19\textheight]{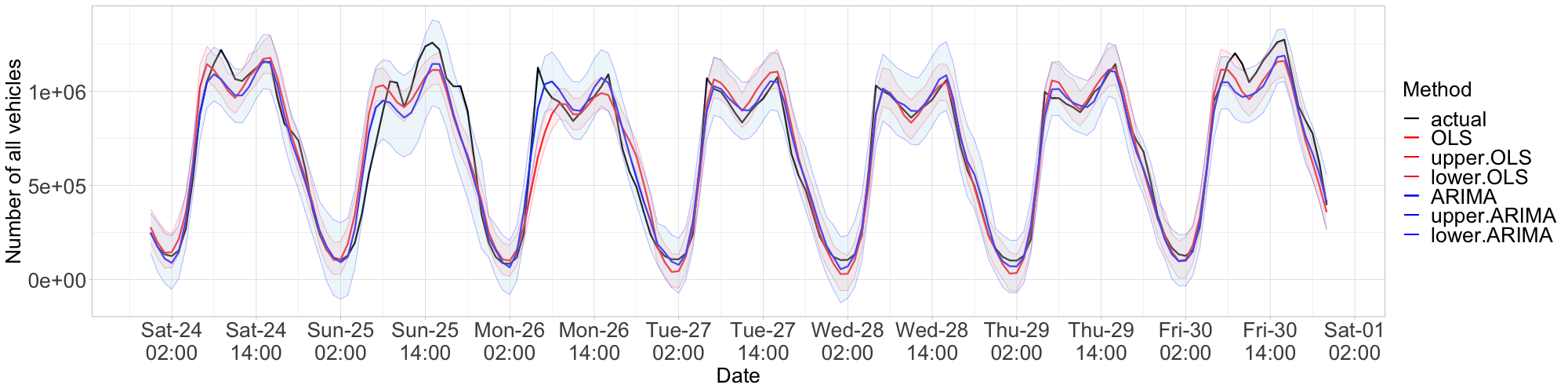}}\\
		\centering
		\subfloat[Northern region, Zhongli runway - Hukou, south direction, car ]{\label{Bottom1}\includegraphics[width=0.95\textwidth, height=0.19\textheight]{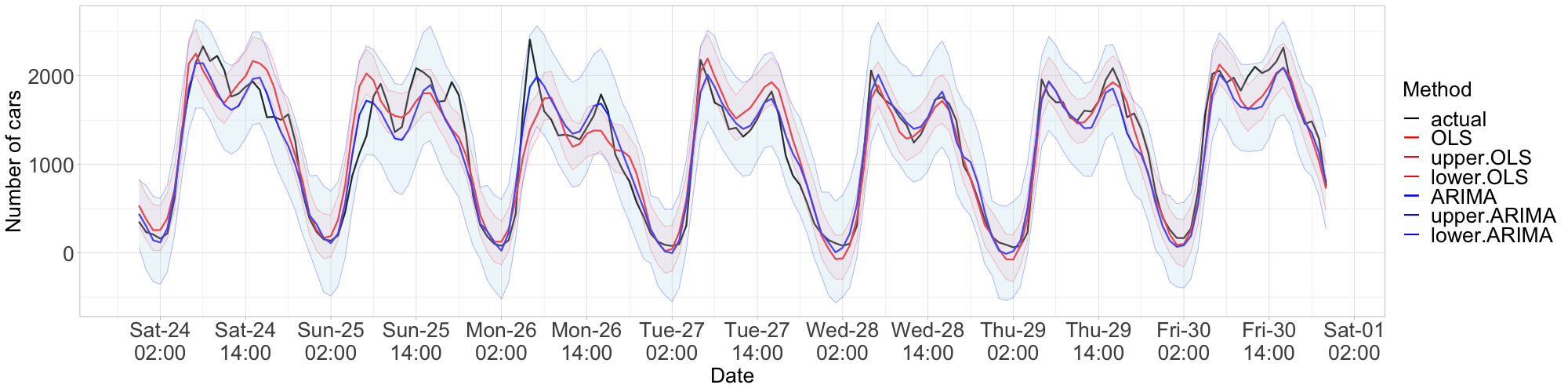}}\\
		\centering
		\subfloat[Central region, Xiangshan - Xibin, south direction, small truck]{\label{Bottom2}\includegraphics[width=0.95\textwidth, height=0.19\textheight]{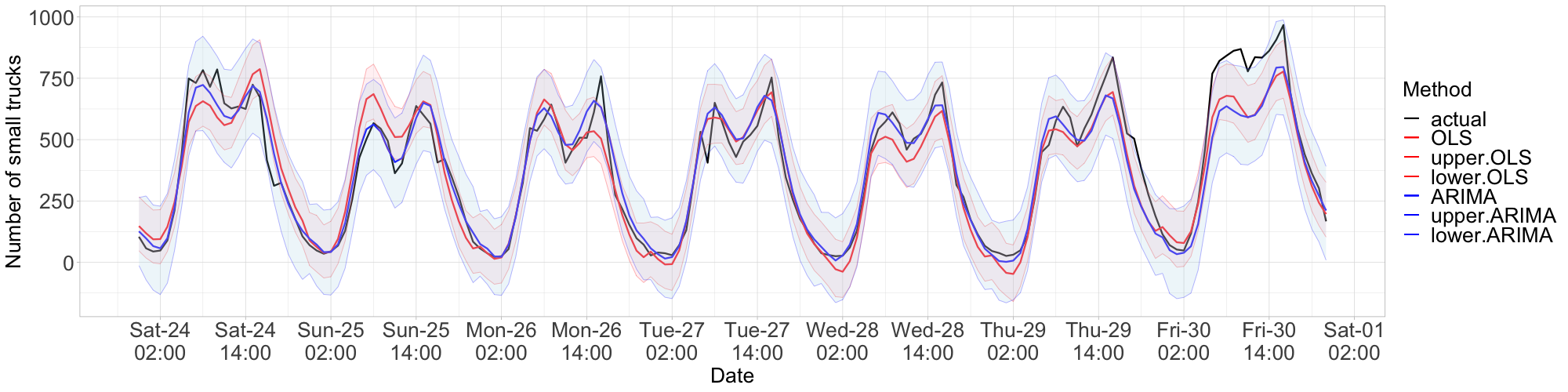}}\\
		\centering
		\subfloat[Southern region, Yanchao SIC - Tianliao, north direction, big truck]{\label{Bottom3}\includegraphics[width=0.95\textwidth, height=0.19\textheight]{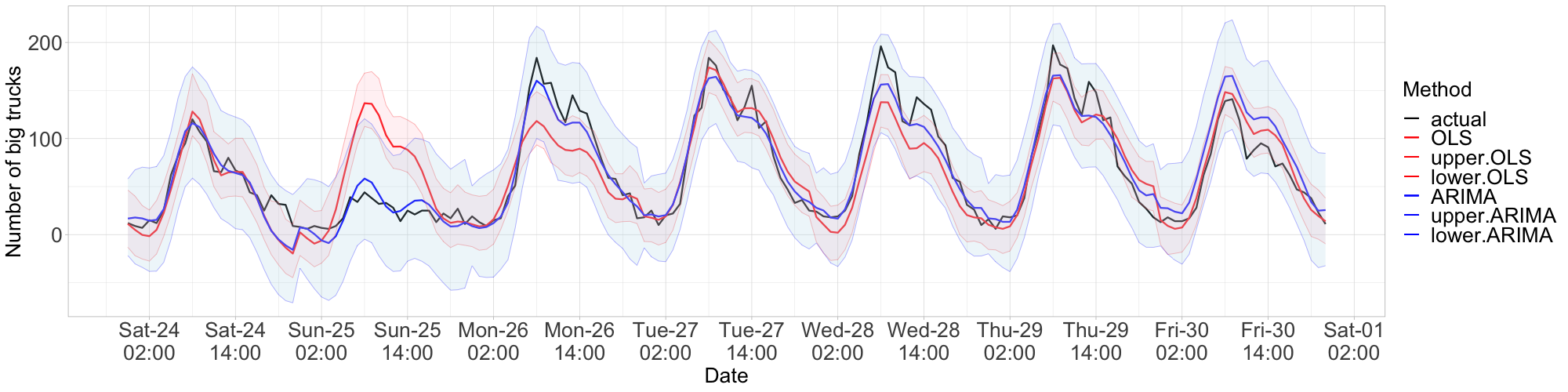}}
		\caption{OLS and ARIMA forecasting with bootstrapped prediction intervals for Taiwan series as well as for the three bottom-level series of Figures \ref{fig:example-Taiwan-traffic-series}-\ref{fig:example-Taiwan-traffic-series-7days} (7 days, from 2021-04-24 to 2021-04-30).}
		\label{fig:compareOLSARIMA-sample series}
	\end{figure}

	Figure \ref{fig:compareOLSARIMA-sample series} compares the OLS and ARIMA approaches' forecast results and prediction intervals for the three bottom-level series of Figures \ref{fig:example-Taiwan-traffic-series}-\ref{fig:example-Taiwan-traffic-series-7days}, as well as for the Taiwan total series, in 7 days (from 2021-04-24 to 2021-04-30). 
	We observe that the forecast results from the two methods are similar in all four series and days, except for the most variable lower count series of Figure \ref{Bottom3} in 2021-04-25. One possible explanation for the fact that the sudden decrease in the number of big trucks traveling on Sundays along this particular highway cannot be forecasted by OLS model, is that it may mainly rely on the previous day pattern, while ARIMA might better capture the pattern that may have occurred during Sundays of previous weeks. 
	In terms of prediction intervals, ARIMA's intervals are generally larger than OLS' ones, especially for the lower count series. 
	Note that the last day depicted in Figure \ref{fig:compareOLSARIMA-sample series} (2021-04-30) is the first day of a consecutive holiday, and in the first three series (Figures \ref{Totalolsarima}, \ref{Bottom1}, and \ref{Bottom2}), the actual traffic flow is higher than the forecasted values for both methods. In particular, both models detect anomalies in Figure \ref{Bottom2}, where they under-forecast the actual series. The big truck series (Figure \ref{Bottom3}) shows a different pattern, in which the observed count is lower than the forecasted one in both OLS and ARIMA approaches.

	Finally, Table \ref{tab:computationtime} compares the computation time of the two approaches, including base forecast, forecast reconciliation, and sample paths (2000 bootstrap samples) reconciliation steps. 
	Note that we forecast each day independently, and this computation time refers to the forecast for one day (one split of the blocked cross-validation). The two methods are run on a Linux server (Compute Canada: \url{docs.alliancecan.ca/wiki/Technical_documentation}) using R version 4.2.1.
	Compared to the ARIMA approach, the OLS model is much faster in the base forecast step despite having similar forecasting performance.

	\subsection{Consecutive holidays anomaly detection}\label{AnomalyResult}
	
	\begin{figure}[!htp]
		\centering
		\subfloat[2019-02-02 to 2019-02-10 - Taiwan]{\label{H1.Total}\includegraphics[width=1.05\textwidth]{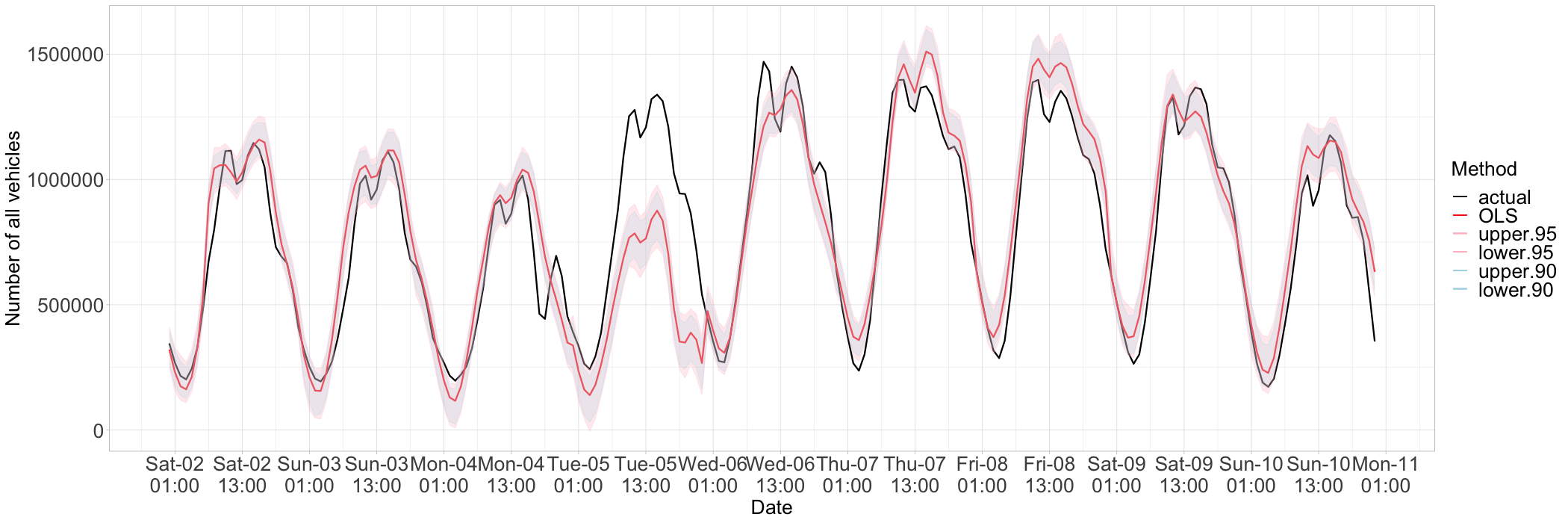}}\\
		\centering
		\subfloat[2020-01-23 to 2020-01-29 - Taiwan]{\label{H7.Total}\includegraphics[width=1.05\textwidth]{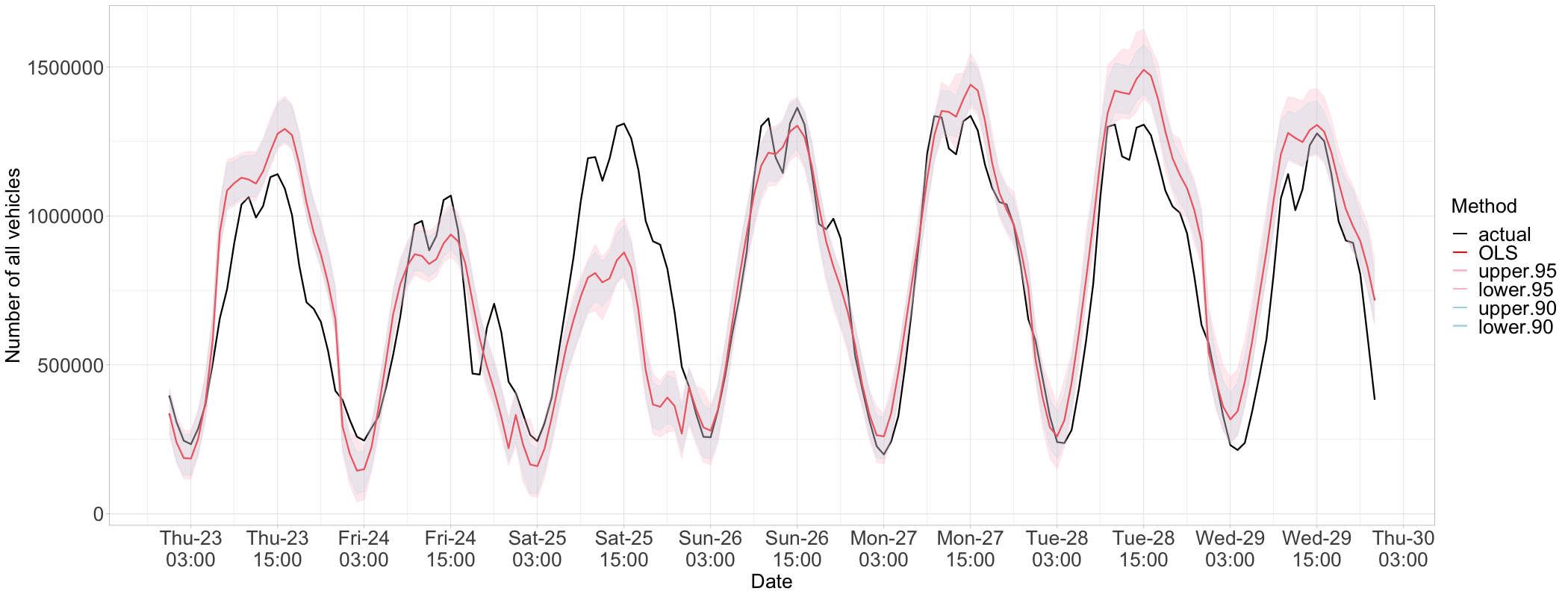}}\\
		\centering
		\subfloat[2021-02-10 to 2021-02-16 - Taiwan]{\label{H15.Total}\includegraphics[width=1.05\textwidth]{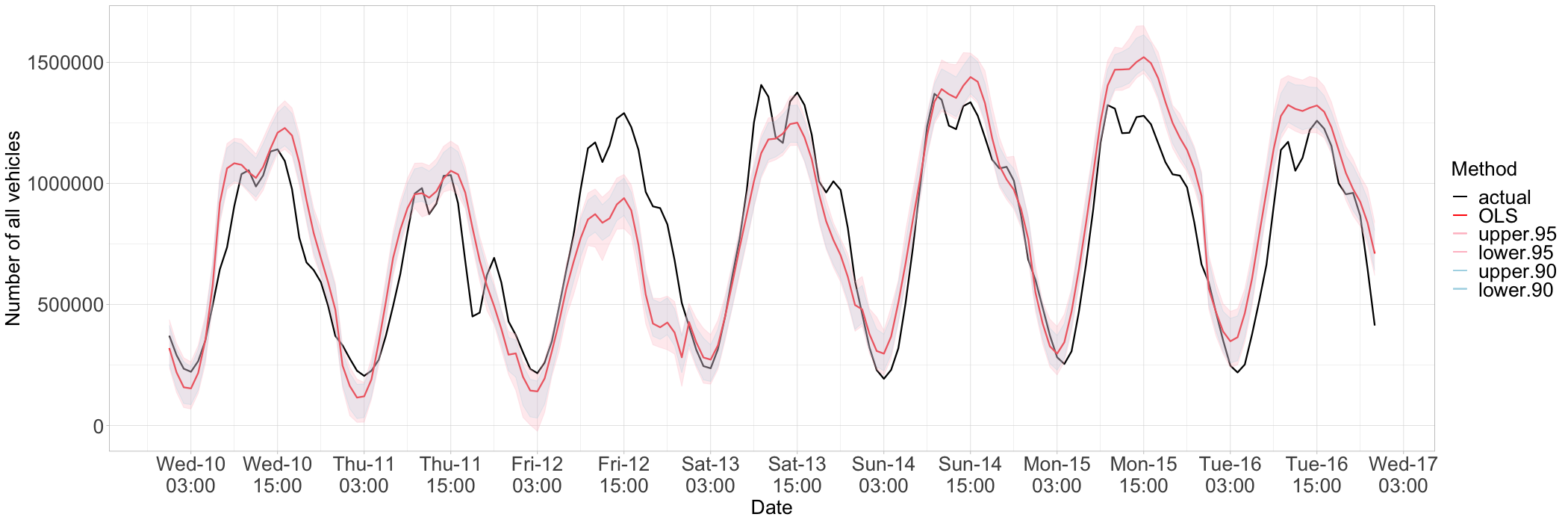}}
		\caption{OLS forecasting with its 90\% and 95\% bootstrapped prediction intervals for Taiwan series in three consecutive long holidays -- 2019-02-02 to 2019-02-10, 2020-01-23 to 2020-01-29, and 2021-02-10 to 2021-02-16.}
		\label{fig:Consecutive-holidays-examples-total}
	\end{figure}
	
	\begin{figure}[!htp]
		\centering
		\subfloat[2019-02-02 to 2019-02-10 - Northern region, Zhongli runway - Hukou, south direction, car]{\label{H1.N2971S31}\includegraphics[width=1.05\textwidth]{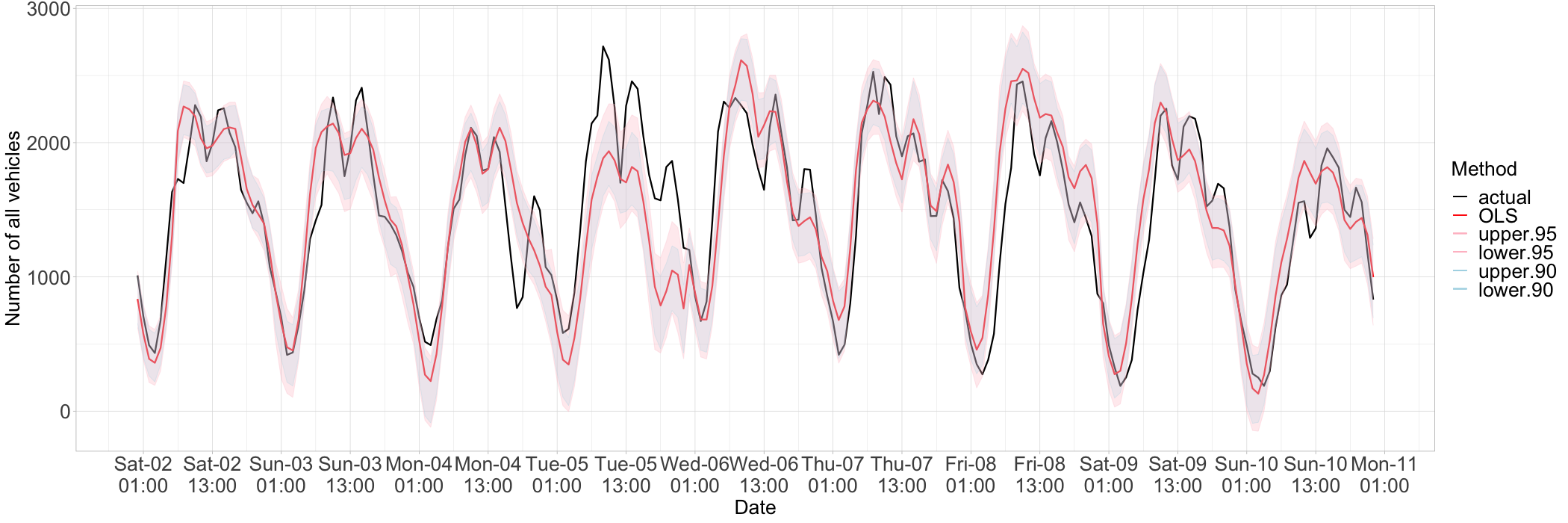}}\\
		\centering
		\subfloat[2020-01-23 to 2020-01-29 - Northern region, Zhongli runway - Hukou, south direction, car]{\label{H7.N2971S31}\includegraphics[width=1.05\textwidth]{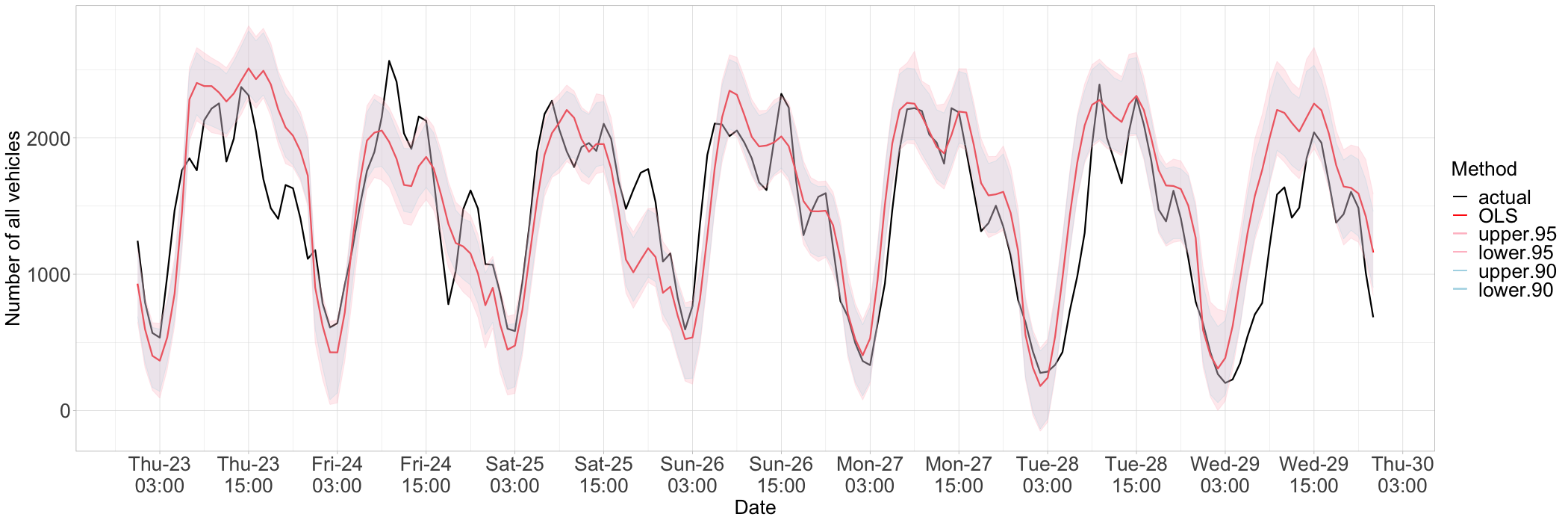}}\\
		\centering
		\subfloat[2021-02-10 to 2021-02-16 - Northern region, Zhongli runway - Hukou, south direction, car]{\label{H15.N2971S31}\includegraphics[width=1.05\textwidth]{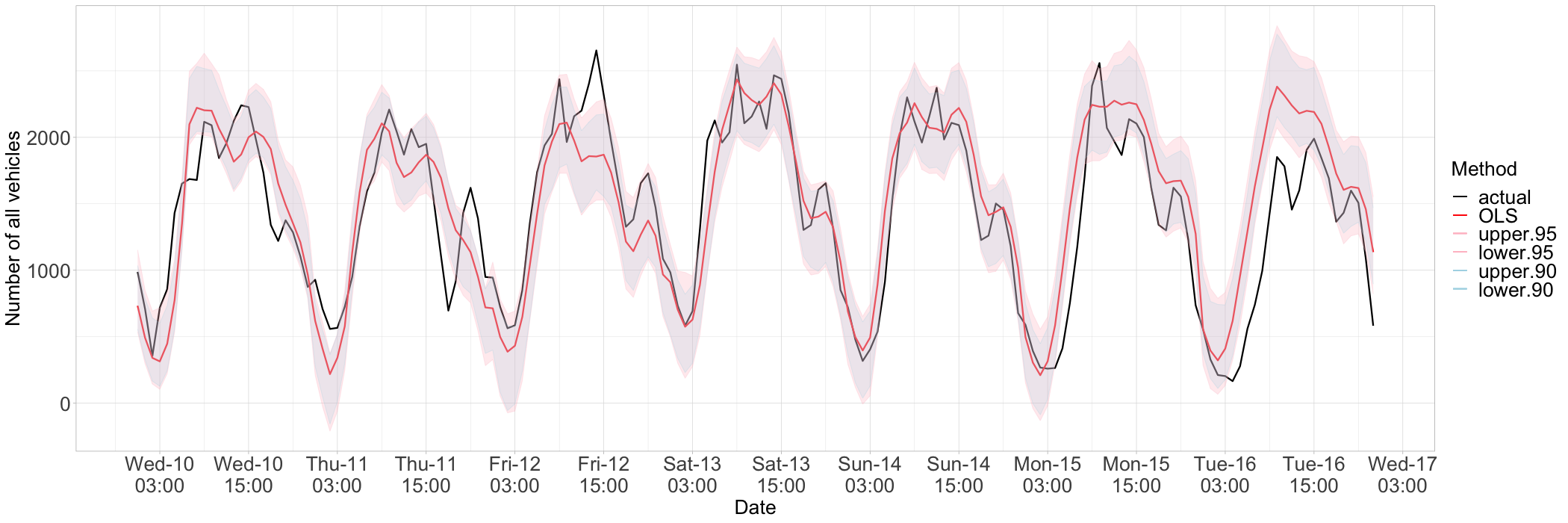}}
		\caption{OLS forecasting with its 90\% and 95\% bootstrapped prediction intervals for the bottom-level series of Figures \ref{N2971S31} and \ref{N2971S31-7days} in three consecutive long holidays -- 2019-02-02 to 2019-02-10, 2020-01-23 to 2020-01-29, and 2021-02-10 to 2021-02-16.}
		\label{fig:Consecutive-holidays-examples}
	\end{figure}
	
	We study the 73 days that are part of 18 consecutive holidays in 2019, 2020, and 2021 (see Table \ref{tab:TCH}), modeling them using the proposed OLS approach and taking 2000 bootstrap samples to generate prediction intervals.  In order to forecast the traffic in each 24-hour period ($T_{test}$; one day) during consecutive holidays, we train the model on the data from the preceding 336 hours ($T_{train}$; two weeks).
    
	Figures \ref{fig:Consecutive-holidays-examples-total} and \ref{fig:Consecutive-holidays-examples} display the OLS forecasting as well as its 90\% and 95\% bootstrapped prediction intervals, for Taiwan total series and for the bottom-level series of Figures \ref{N2971S31} and \ref{N2971S31-7days}, in the three longest consecutive holidays, i.e.~2019-02-02 to 2019-02-10, 2020-01-23 to 2020-01-29, and 2021-02-10 to 2021-02-16. 
\begin{table}[!tb]
\footnotesize
\caption{\label{tab:anomalies-region-by-directionNS} 
			The number of detected anomalies per day (divided by holiday length) in the 18 long holidays in the northern, central, and southern regions in the north and south, directions, divided into four time intervals: 00:00-05-00 (1), 06:00-11:00 (2), 12:00-17:00 (3), and 18:00-23:00 (4). Cell background colors are proportional to the number of anomalies per day (white for 0-1.5 anomalies, light gray for 1.5-3 anomalies, gray for 3-4.5 anomalies, and dark gray for 4.5 or more anomalies). 
		}
    \centering
    \begin{tabular}{lrrrrrrrrrrrr}
     \toprule
			\multicolumn{1}{c}{} & \multicolumn{12}{c}{North direction} \\
			\multicolumn{1}{c}{} & \multicolumn{4}{c}{North} & \multicolumn{4}{c}{Center} & \multicolumn{4}{c}{South}\\
			\cmidrule(l{3pt}r{3pt}){2-5} \cmidrule(l{3pt}r{3pt}){6-9} \cmidrule(l{3pt}r{3pt}){10-13} 
			Holiday &1 & 2 & 3 & 4 &1 & 2 & 3 & 4 &1 & 2 & 3 & 4 \\
			\midrule
			2019&&&&&&&&&&&&\\
			02-02 to 02-10 &0.0&0.5&0.6&1.2&0.0&1.3&1.0&1.2&0.0&1.4&0.8&0.7\\
			02-28 to 03-03 &0.0&0.7&1.5&\cellcolor{gray!15}2.5&0.0&1.5&\cellcolor{gray!15}3.0&\cellcolor{gray!15}1.7&0.0&1.5&\cellcolor{gray!15}1.7&\cellcolor{gray!15}1.7\\
			04-04 to 04-07 
             & 0.0& 1.0& \cellcolor{gray!15}2.7& \cellcolor{gray!15}3.0& 0.0& 1.2 & \cellcolor{gray!15}2.7& \cellcolor{gray!15}2.7& 0.0&\cellcolor{gray!15} 3.0& \cellcolor{gray!15}3.0 & \cellcolor{gray!15}3.0\\
			06-07 to 06-09 
            &0.0 &1.3& \cellcolor{gray!15}2.0& \cellcolor{gray!15}2.0& 0.0& \cellcolor{gray!15}1.6& \cellcolor{gray!40}3.3& \cellcolor{gray!15}2.3& 0.0& \cellcolor{gray!15}1.6& \cellcolor{gray!15}2.6&1.3\\
			09-13 to 09-15 
            &0.0& 0.6&\cellcolor{gray!15} 2.0&\cellcolor{gray!15} 2.0& 0.0& \cellcolor{gray!15}2.0& \cellcolor{gray!40}3.3& \cellcolor{gray!40}3.3& 0.0& \cellcolor{gray!15}2.3& \cellcolor{gray!40}3.3& \cellcolor{gray!15}2.3\\
			10-10 to 10-13 
            &0.0& 0.7& 1.5&\cellcolor{gray!15} 2.5& 0.0& \cellcolor{gray!15}2.0& \cellcolor{gray!15}2.7& \cellcolor{gray!15}2.2& 0.0&\cellcolor{gray!15} 2.2& \cellcolor{gray!15}2.7&\cellcolor{gray!15} 2.2\\
			2020&&&&&&&&&&&&\\
   			01-23 to 01-29 
             &0.7 &0.7& 1.1& \cellcolor{gray!15}2.0& 0.4&\cellcolor{gray!15} 2.0& \cellcolor{gray!15}2.0& \cellcolor{gray!15}2.8& 0.4&\cellcolor{gray!15} 2.0& \cellcolor{gray!15}1.8& \cellcolor{gray!15}1.8\\
			02-28 to 03-01 
             &0.0 &0.6& \cellcolor{gray!15}2.0& \cellcolor{gray!15}2.6& 0.0&\cellcolor{gray!15} 2.0& \cellcolor{gray!40}4.0& \cellcolor{gray!40}3.3& 0.0&\cellcolor{gray!15} 2.6& \cellcolor{gray!60}4.6& \cellcolor{gray!15}3.0\\
			04-02 to 04-05 
            &0.0& 0.7& \cellcolor{gray!15}2.2& \cellcolor{gray!15}2.7& 0.0& 1.0& \cellcolor{gray!40}3.5&\cellcolor{gray!15} 2.7& 0.0&0.7& \cellcolor{gray!15}3.0& \cellcolor{gray!15}2.7\\
		05-01 to 05-03 
            &0.0& 0.6& \cellcolor{gray!15}2.0 &\cellcolor{gray!15}3.0& 0.0& \cellcolor{gray!15}2.3& \cellcolor{gray!40}4.0& \cellcolor{gray!40}3.6& 0.0& 1.6& \cellcolor{gray!40}3.6& \cellcolor{gray!40}3.3\\
			06-25 to 06-28  
            &0.0& 1.0& \cellcolor{gray!15}1.7& \cellcolor{gray!15}2.5& 0.0&\cellcolor{gray!15} 2.0& \cellcolor{gray!15}2.5&\cellcolor{gray!15}2.2& 0.7& \cellcolor{gray!15}2.7& \cellcolor{gray!40}3.2& \cellcolor{gray!15}2.2\\
			10-01 to 10-04 
            &0.0 &0.5& 1.5&\cellcolor{gray!15} 2.2& 0.0& 1.5& \cellcolor{gray!15}2.5& \cellcolor{gray!15}2.2& 0.0& 1.5& \cellcolor{gray!15}2.7& \cellcolor{gray!15}2.5\\
		    10-09 to 10-11 
            &0.0& 0.0& 0.0&0.3& 0.0& 1.0& \cellcolor{gray!15}2.0& 1.3& 0.0&1.0& 1.3& 1.3 \\
            2021&&&&&&&&&&&&\\
		01-01 to 01-03  
            &0.6& 0.6& \cellcolor{gray!15}2.0& \cellcolor{gray!15}2.0& 0.0& 1.0& \cellcolor{gray!15}3.0& \cellcolor{gray!15}2.6& 0.0& 1.6& \cellcolor{gray!15}2.6& \cellcolor{gray!15}2.6\\
			02-10 to 02-16  
           &0.2& 0.7& 1.0& \cellcolor{gray!15}2.2& 0.0& 1.2& 1.5& \cellcolor{gray!15}2.8& 0.0&1.1& 1.4& \cellcolor{gray!15}2.5\\
			02-27 to 03-01 
           &0.0& 1.3& 1.3& 1.3 &0.0& 1.3& \cellcolor{gray!15}2.3& 1.3& 0.0& \cellcolor{gray!15}1.6& 1.0& 1.3\\
			04-02 to 04-05 
           & 0.5 &0.5 &1.5 &\cellcolor{gray!15}2.2& 0.0& 1.0& \cellcolor{gray!40}3.7& \cellcolor{gray!15}2.5& 0.0& 1.2&  \cellcolor{gray!15}3.0& \cellcolor{gray!15}1.7\\
		04-30 to 04-30 
           &0.0& 0.0&0.0& 0.0& 0.0& 0.0& 0.0& 0.0& 0.0&\cellcolor{gray!15} 2.0& 1.0&0.0\\
           \midrule
           \multicolumn{1}{c}{} & \multicolumn{12}{c}{South direction} \\
			\multicolumn{1}{c}{} & \multicolumn{4}{c}{North} & \multicolumn{4}{c}{Center} & \multicolumn{4}{c}{South}\\
			\cmidrule(l{3pt}r{3pt}){2-5} \cmidrule(l{3pt}r{3pt}){6-9} \cmidrule(l{3pt}r{3pt}){10-13} 
			Holiday &1 & 2 & 3 & 4 &1 & 2 & 3 & 4 &1 & 2 & 3 & 4 \\
			\midrule
			2019&&&&&&&&&&&&\\
			02-02 to 02-10 &0.1&0.3&0.6&1.1&0.7&1.5&1.3&0.8&0.3&1.2&0.7&0.6\\
			02-28 to 03-03 &0.0&1.0&1.5&\cellcolor{gray!15}2.2&0.0&\cellcolor{gray!15}2.8&\cellcolor{gray!15}1.7&0.0&0.0&\cellcolor{gray!15}2.2&\cellcolor{gray!15}2.5&0.7\\
			04-04 to 04-07 
             & 0.2& 1.0& \cellcolor{gray!15}2.2& \cellcolor{gray!15}2.7& 0.2&\cellcolor{gray!15} 2.5&1.2& 0.2& 0.0& 1.2& \cellcolor{gray!15}2.5& \cellcolor{gray!15}2.5\\
			06-07 to 06-09 
            & 0.0& 0.6& 0.6& 0.6& 0.6& \cellcolor{gray!15}2.0& \cellcolor{gray!40}3.3& \cellcolor{gray!40}3.3& 0.0& 1.3& \cellcolor{gray!40}3.3& \cellcolor{gray!40}3.3\\    
			09-13 to 09-15 
            & 0.0& 0.6& \cellcolor{gray!15}2.0& \cellcolor{gray!15}2.0& 0.0& 0.6& 1.3& 1.0& 0.0& 0.6& 1.3& \cellcolor{gray!15}1.6\\    
			10-10 to 10-13 
            & 0.2& 1.0& 1.5& \cellcolor{gray!15}2.0& 0.0&\cellcolor{gray!15} 2.5& \cellcolor{gray!15}2.0& 1.0& 0.0& \cellcolor{gray!15}2.2& \cellcolor{gray!40}3.0& \cellcolor{gray!15}2.2\\ 
			2020&&&&&&&&&&&&\\
   			01-23 to 01-29 
             & 0.5& 0.1&1.4& 1.4& 0.4& \cellcolor{gray!15}1.8& \cellcolor{gray!15}1.7& 0.8& 0.5& 1.2& 0.8& 1.2\\
			02-28 to 03-01 
             & 0.0& 1.0&\cellcolor{gray!15}2.3& \cellcolor{gray!15}2.6& 0.0& 1.3& \cellcolor{gray!40}4.0& 1.3& 0.0& \cellcolor{gray!15}1.6& \cellcolor{gray!40}3.3& \cellcolor{gray!15}2.6\\
			04-02 to 04-05 
            & 0.0& 0.7& 1.5& \cellcolor{gray!15}2.7& 0.0& 0.7& \cellcolor{gray!15}2.0& 0.5& 0.0&0.5& \cellcolor{gray!40}3.2& \cellcolor{gray!15}2.0\\
		05-01 to 05-03 
             & 0.0& 0.6 &\cellcolor{gray!15}2.0& \cellcolor{gray!15}2.6& 0.0& \cellcolor{gray!15}2.0& \cellcolor{gray!40}4.3& \cellcolor{gray!15}1.6& 0.0& 1.0& \cellcolor{gray!40}3.3& \cellcolor{gray!15}3.0\\
			06-25 to 06-28  
            & 0.2& 0.7& 1.5& \cellcolor{gray!15}2.2& 0.0& \cellcolor{gray!15}1.7& 1.5& 0.0& 0.0& 1.0& \cellcolor{gray!15}2.5& 1.5\\
			10-01 to 10-04 
           & 0.5& 0.5& 1.5& \cellcolor{gray!15}1.7& 0.0& 1.0& 1.2& 0.2& 0.0& 0.5& \cellcolor{gray!15}1.7& 1.5\\
		    10-09 to 10-11 
            &0.0& 0.0& 0.0& 0.0& 0.0& \cellcolor{gray!40}3.3&\cellcolor{gray!15} 2.3& \cellcolor{gray!15}1.6& 0.0& \cellcolor{gray!15}1.6& \cellcolor{gray!15}1.6& 1.3\\
            2021&&&&&&&&&&&&\\
		01-01 to 01-03  
            & 0.6& 1.0& \cellcolor{gray!15}2.0& \cellcolor{gray!15}2.0& 0.0& 0.6&1.0& 1.3& 0.0& 0.6& \cellcolor{gray!15}2.3& \cellcolor{gray!15}3.0\\
			02-10 to 02-16  
           & 0.0& 0.1& 0.8& \cellcolor{gray!15}2.1& 0.2& 1.4& 1.0& 0.8& 0.0& 0.8&0.8& 1.4\\
			02-27 to 03-01 
           & 0.0& 1.0& 0.6 &1.0& 0.0&  \cellcolor{gray!15}& 1.0& 0.3& 0.0& 0.6& 1.3& \cellcolor{gray!15}1.6\\
			04-02 to 04-05 
           & 0.2& 1.5& 1.0&\cellcolor{gray!15} 2.0& 0.0& \cellcolor{gray!15}1.7&\cellcolor{gray!15} 1.7& 0.2& 0.0& 0.5&  \cellcolor{gray!15}2.0& 1.0\\
		04-30 to 04-30 
           & 0.0& 0.0& 0.0& 0.0& \cellcolor{gray!40}4.0& 0.0& 0.0& 0.0&\cellcolor{gray!15} 2.0& 0.0&0.0& 0.0\\
			\bottomrule
		\end{tabular}
\end{table}

 \begin{table}[!tb]
\footnotesize
\caption{\label{tab:anomalies-region-by-directionEW} 
			The number of detected anomalies per day (divided by holiday length) in the 18 long holidays in the northern region in the east and west, directions, divided into four time intervals: 00:00-05-00 (1), 06:00-11:00 (2), 12:00-17:00 (3), and 18:00-23:00 (4). Cell background colors are proportional to the number of anomalies per day (white for 0-1.5 anomalies, light gray for 1.5-3 anomalies, gray for 3-4.5 anomalies, and dark gray for 4.5 or more anomalies). 
		}
		\centering
		\begin{tabular}[ht]{lrrrrrrrr}
			\toprule
			\multicolumn{1}{c}{} & \multicolumn{4}{c}{East direction} &  \multicolumn{4}{c}{West direction}\\
			\cmidrule(l{3pt}r{3pt}){2-5} \cmidrule(l{3pt}r{3pt}){6-9} 
			\multicolumn{1}{c}{} & \multicolumn{4}{c}{North} & \multicolumn{4}{c}{North}\\
			\cmidrule(l{3pt}r{3pt}){2-5} \cmidrule(l{3pt}r{3pt}){6-9}  
			Holiday &1 & 2 & 3 & 4 &1 & 2 & 3 & 4\\
			\midrule
			2019&&&&&&&&\\
			02-02 to 02-10 &0.0&0.4&0.6&1.0&0.0&0.2&0.6&0.4\\
			02-28 to 03-03 &0.0&1.2&0.8&1.0&0.0&0.5&1.2&\cellcolor{gray!15}2.3\\
			04-04 to 04-07 
             & 0.0& \cellcolor{gray!15}1.2&\cellcolor{gray!15} 2.0&\cellcolor{gray!15} 2.7& 0.0& 0.5& \cellcolor{gray!15}3.0& \cellcolor{gray!15}2.7\\
			06-07 to 06-09 
            & 1.3& \cellcolor{gray!15}2.0& 1.3& 0.0& 0.0& 0.6& \cellcolor{gray!15}2.0& \cellcolor{gray!15}2.0\\
			09-13 to 09-15 
            & 0.0& 0.6& \cellcolor{gray!15}2.0& \cellcolor{gray!15}2.0& 0.0& 0.6& \cellcolor{gray!15}2.0&\cellcolor{gray!15}2.3\\
			10-10 to 10-13 
          & 0.0& 1.0& 1.5& \cellcolor{gray!15}1.7& 0.0& 0.7& 1.5& 1.0\\
			2020&&&&&&&&\\
   			01-23 to 01-29 
             & 0.0& 0.4&0.7& 1.1& 0.0&0.2&\cellcolor{gray!15}1.7& 1.1\\
			02-28 to 03-01 
             & 0.0& 0.6& \cellcolor{gray!15}1.6& 1.3& 0.0&0.6&\cellcolor{gray!15}2.0& 2.0\\
			04-02 to 04-05 
            & 0.0& 0.5& 1.2& \cellcolor{gray!15}1.7& 0.0& 0.5& 1.5& \cellcolor{gray!15}2.2\\
		05-01 to 05-03 
           & 0.0& 0.6& 1.0&0.6& 0.0& 0.6&\cellcolor{gray!15} 2.0& \cellcolor{gray!15}2.0\\
			06-25 to 06-28  
           & 0.0& 0.7& 0.7& \cellcolor{gray!15}1.7& 0.0& 0.5& 1.5&1.0\\
			10-01 to 10-04 
           & 0.0& 0.7& 1.5&\cellcolor{gray!15} 2.0& 0.0& 0.2 &1.5& 1.2\\
		    10-09 to 10-11 
             & 0.0& 0.0& 0.0& 0.0& 0.0& 0.0&0.0& 0.0\\
            2021&&&&&&&&\\
		01-01 to 01-03  
            & 1.0& 1.0& \cellcolor{gray!15}2.0& \cellcolor{gray!15}1.6& 0.0& 1.0& \cellcolor{gray!15}2.0& \cellcolor{gray!15}2.0\\
			02-10 to 02-16  
           & 0.0& 0.2& 0.4& \cellcolor{gray!15}1.7& 0.0& 0.1& 0.8& 1.5\\
			02-27 to 03-01 
           & 0.0& 1.0& 1.0&0.3& 0.0& 0.6& 0.6& 0.6\\
			04-02 to 04-05 
          & 0.0& 0.7& 1.2& 1.2& 0.0& 0.5& 1.5& 1.5\\
		04-30 to 04-30 
           & 0.0& 0.0& 0.0& 0.0& 0.0& 0.0& 0.0& 0.0\\
			\bottomrule
		\end{tabular}
\end{table}
        \begin{sidewaysfigure}[!htp]
		\centering
		\begin{tikzpicture}
			\node[scale=0.35]
			{ \scalebox{1}[1]{\includegraphics[scale=0.7]{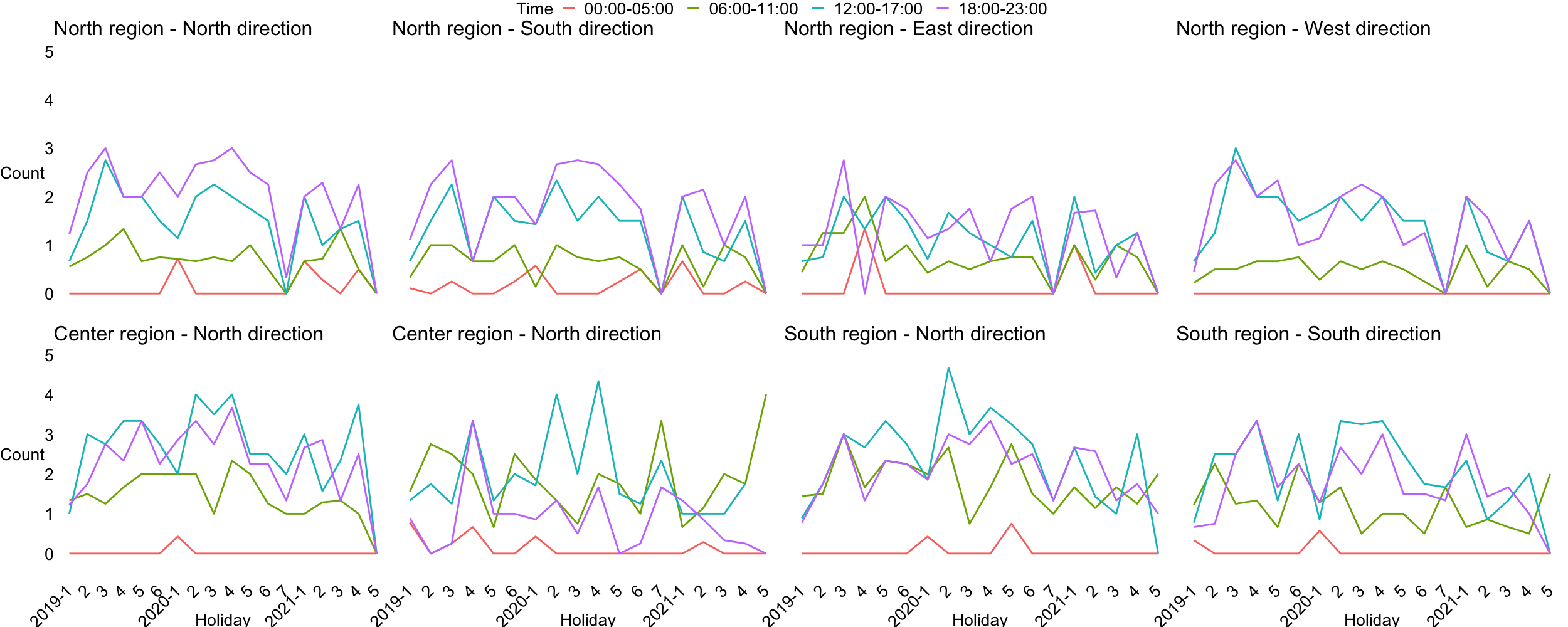}} };  
		\end{tikzpicture}
		\caption{
			Number of detected anomalies per day (divided by holiday length) in the 18 long holidays in the northern, central, and southern regions in the north, south, east, and west directions, divided in four-time intervals: 00:00-05-00 (1), 06:00-11:00 (2), 12:00-17:00 (3), and 18:00-23:00 (4).}
		
		\label{fig:anomalies-region-by-direction}
	    \end{sidewaysfigure}
        Figure \ref{fig:Consecutive-holidays-examples-total} illustrates traffic flow in Taiwan (total series) during the Lunar New Year holidays, highlighting unexpected spikes (anomalies) on the first day of the Chinese New Year (2019-02-05, 2020-01-25, and 2021-02-12).
        These two figures also show that the prediction intervals at 90\% and 95\% are pretty similar and anomaly detection is not very sensitive to changing the prediction level from  90\% to 95\%; thus, we focus on the 95\% prediction intervals in the following results. 
        
	Tables \ref{tab:anomalies-region-by-directionNS} and \ref{tab:anomalies-region-by-directionEW} report the number of detected anomalies per day (divided by holiday length) in each of the three regions and in each direction. In these tables, the day is divided into four time intervals, i.e.~00:00 to 05:00, 06:00 to 11:00, 12:00 to 17:00, and 18:00 to 23:00. 
 
	The same results are also displayed in Figure \ref{fig:anomalies-region-by-direction}. Additional anomaly detection results based on the highway direction and number are presented in Tables \ref{tab:anomalies-highway-by-directionNS},  \ref{tab:anomalies-highway-by-directionEW} and Figure \ref{fig:anomalies-highway-by-direction}. In general, we detect a small number of anomalies during the night (from 00:00 to 05:00), especially in the east and west directions. 
    In the north direction, anomalies are mostly in the center and south regions, in the afternoon and evening (from 12:00 to 23:00), while in the south direction, anomalies are typically detected in the center region, in the morning and afternoon (from 6:00 to 17:00) and in the south region during afternoon and evening hours (from 12:00 to 23:00). 
    Highways No.1 and No.3 have more anomalies in both north and south directions, with No.3 exhibiting a higher anomaly frequency than No.1 on both highways. Anomalies on Highway No.1 in the north direction are more frequent in the evening hours (from 18:00 to 23:00), while on Highway No.3, anomalies occur more frequently in the afternoon and evening (from 12:00 to 23:00). In the south direction, the No.1 highway exhibits more anomalies from 12:00 to 23:00, while the No.3 highway has a higher anomaly frequency from 12:00 to 18:00.
    
	In the west direction of the elevated No.1 highway, there are similar portions of anomalies during afternoons and evenings (from 12:00 to 17:00 and from 18:00 to 23:00). 
    Interestingly, the number of anomalies has remained relatively stable from 2019 to 2021, with no discernible upward or downward trend (Figure \ref{fig:anomalies-region-by-direction}).

    \begin{table}[!htp]
		\caption{\label{tab:anomsummerize} Summary of the most frequent anomalies on consecutive holidays by direction, region, highway, and time interval in 2019, 2020, and 2021.}
		\centering
		\begin{tabular}[t]{llllll}
			\toprule
			Year & &  North direction & South direction & East direction & West direction \\
			\midrule
			2019 & Region & Center/South & South & North & North\\
			& Highway & No.1/No.3 & No.3 & Elevated No.1& Elevated No.1\\
			& Time &12:00-23:00 &12:00-23:00& 18:00-23:00 & 12:00-23:00\\
			\midrule
			2020 & Region & Center/South & South & North & North\\
			& Highway & No.1/No.3 & No.3 & Elevated No.1 & Elevated No.1\\
			& Time & 12:00-23:00 & 12:00-17:00 & 18:00-23:00 & 12:00-23:00\\
			\midrule
			2021 & Region & Center/South & Center & North & North\\
			& Highway & No.1/No.3 & No.3 & Elevated No.1 & Elevated No.1\\
			& Time & 12:00-23:00 & 6:00-17:00 & 18:00-23:00 & 12:00-23:00\\
			\bottomrule
		\end{tabular}
	\end{table}
    \begin{table}[!hbp]
		\caption{\label{tab:Similarity}The percentage of similarity in anomaly detection results (i.e.,~proportion of time steps which are classified in the same way) between OLS and ARIMA with reconciliation step (columns Rec-ARIMA) and without reconciliation step (columns ARIMA) during consecutive holidays in Taiwan from 2019-01-01 to 2021-04-30.}
		\centering
		\begin{tabular}[t]{lrrrrr}
  \multicolumn{1}{c}{} & \multicolumn{1}{c}{} & \multicolumn{2}{c}{Rec-ARIMA}&
  \multicolumn{2}{c}{ARIMA}\\
			\toprule
			Year & Consecutive holidays & Region & Highway & Region & Highway \\
			\midrule
			2019 & \\
			& 02-02 to 02-10 (9 days)& 89\% & 89\% & 74\% & 76\%\\
			& 02-28 to 03-03 (4 days)& 84\% & 85\% & 68\% & 66\%\\
			& 04-04 to 04-07 (4 days)& 72\% & 75\% & 64\% & 62\%\\
			& 06-07 to 06-09 (3 days)& 71\% & 75\% & 65\% & 66\%\\
			& 09-13 to 09-15 (3 days)& 73\% & 76\% & 67\% & 67\%\\
			& 10-10 to 10-13 (4 days)& 79\% & 79\% & 65\% & 64\%\\
			\hline
			2020 & \\
			& 01-23 to 01-29 (7 days)& 78\% & 80\% & 68\% & 69\%\\
			& 02-28 to 03-01 (3 days)& 83\% & 81\% & 57\% & 59\%\\
			& 04-02 to 04-05 (4 days)& 78\% & 79\% & 64\% & 66\%\\
			& 05-01 to 05-03 (3 days)& 78\% & 80\% & 58\% & 61\%\\
			& 06-25 to 06-28 (4 days)& 76\% & 77\% & 65\% & 66\%\\
			& 10-01 to 10-04 (4 days)& 79\% & 81\% & 67\% & 69\%\\
			& 10-09 to 10-11 (3 days)& 91\% & 93\% & 73\% & 77\%\\
			\hline
			2021 & \\
			& 01-01 to 01-03 (3 days)& 75\% & 77\% & 64\% & 63\%\\
			& 02-10 to 02-16 (7 days)& 83\% & 83\% & 70\% & 71\%\\
			& 02-27 to 03-01 (3 days)& 86\% & 88\% & 69\% & 71\%\\
			& 04-02 to 04-05 (4 days)& 79\% & 79\% & 69\% & 68\%\\
			& 04-30 to 04-30 (1 day)& 91\% & 95\% & 75\% & 78\%\\
			\hline
		\end{tabular}
	\end{table}
 
     Table \ref{tab:anomsummerize} summarizes the most frequent traffic flow anomalies in Taiwan's consecutive holidays in 2019, 2020, and 2021. 
	We observe that in the north direction, anomalies are mainly in the central and southern areas of highways No.1 and No.3 during afternoons and evenings (from 12:00 to 23:00). 
	In the south direction, most traffic flow anomalies are in the southern region of Highway No.3 during the afternoon and evening (from 12:00 to 17:00) in 2019 and 2020, while in 2021, anomalies are mostly in the central area during the morning hours (from 6:00 to 17:00). 
	The elevated No.1 highway is the only highway that includes the east and west directions (in the northern region). In this highway, anomalies in the east direction are mostly detected during the evening (from 18:00 to 23:00), while in the west direction, anomalies are mostly detected during the afternoon and evening (from 12:00 to 23:00).
 

Finally, Table \ref{tab:Similarity} presents a comparative analysis of anomaly detection results. In particular, the columns Rec-ARIMA present the percentage of similarity in anomaly detection results between our proposed OLS approach and the ARIMA reconciled method (see Section \ref{sec:forecastingResult}). The columns ARIMA contrast the OLS approach results with the ARIMA method without reconciliation step, which is similar to the time series outlier detection method. This evaluation includes the series in diverse directions within various regions and highways in all 18 consecutive holidays. This table reveals that the anomalies identified by the OLS approach align well with those detected by the reconciled ARIMA technique. In contrast, the similarity is less apparent when comparing our approach with the ARIMA model without reconciliation. This is consistent with our expectations, as the unreconciled ARIMA approach lacks the capability to capture the spatial correlation between the traffic flow series.

\section{Conclusion}\label{conclusion}
	
	In this study, we explored the Taiwanese highway hourly traffic data with the main objective of detecting anomalies during long holidays (e.g.,~Chinese New Year). 
	Understanding where and when most anomalies occur is of utmost importance to help the Taiwanese government and the National Freeway Bureau evaluate and adjust their traffic control policies. 
	Our results suggest that, based on the highway direction, the north and south directions show more anomalies on highway No.3 in the afternoon, central and southern Taiwan in the north direction, and southern Taiwan in the south. 
	East and west directions included fewer anomalies, which were mainly in the 12:00-23:00 period. 
	
	To perform this traffic anomaly detection, we employed a fast OLS approach, which allows one to efficiently model the complex hourly traffic dataset with reasonable accuracy and bootstrapped prediction intervals, which rely only on the assumption of uncorrelated forecast errors. 
	The proposed model captures seasonality and spatial correlation of traffic using Fourier terms and hierarchical aggregation. While other more complex models like ARIMA require high computation time for forecasting complex datasets, our OLS model is computationally easy to handle. 
	This computational efficiency is particularly useful when utilizing a blocked cross-validation scheme to forecast long time series since this approach requires fitting the model and forecasting many times (once for each cross-validation split). 
	We also observe that, while the proposed method may not be considered a real-time prediction method, it nevertheless generates accurate forecasts quickly on a daily basis, and it could be easily adapted to generate forecasts more frequently (for example, every hour). 
	In addition to its computational efficiency, other advantages of the proposed OLS model over more complex approaches are its flexibility in adding external information (e.g.,~dummy variables to capture holiday information or different traffic management strategies) and its interpretability. 
	Note that in this study, we did not consider external information, such as holidays, directly in the model because of our goal of detecting traffic anomalies. 
	We observe that, although the results presented in this paper concern a static model (i.e.,~the same predictors for all cross-validation splits), the model could be updated and adapted to capture the latest changes in time series patterns. 
	Our proposed model also has the practical advantage of handling missing values by simply automatically removing them, while models like ARIMA require imputation. 

The application of OLS models in forecasting and anomaly detection can raise concerns regarding result robustness. In our proposed approach, switching from OLS to more resilient methods like robust regression is straightforward, although it may result in a slight increase in computation time. We run our method using robust regression instead of OLS, and found that the results did not significantly differ. Hence, we presented only the OLS version of our method because it offered similar results in our application, while being computationally more efficient.
	
	Earlier detection of unusual traffic events is critical for traffic authorities' decision-making to maintain smooth mobility. 
	The results of this study can be easily updated daily (or hourly), which can be very valuable to the national freeway bureau for preventing traffic jams and congestion, as well as for improving management policies.  While we applied the OLS model for forecasting and anomaly detection on Taiwanese highways, this model can also be applied to areas with more complex road structures and a higher number of road divisions. In fact, the OLS model remains computationally efficient even with an increased number of time series, whereas models like ARIMA experience linear increases in computation time as the number of series grows \citep{ashouri2021fast}. 


\section*{Acknowledgements}
 Frederick Kin Hing Phoa acknowledges the support of the Academia Sinica (Taiwan) grant number AS-IA-112-M03, and the National and Science Council (Taiwan) grant number 111-2118-M-001-007-MY2. 
 M.A.~Cremona acknowledges the support of the Natural Sciences and Engineering Research Council of Canada (NSERC) grant number RGPIN-2020-05657, and of FSA, Université Laval.

\section*{Supplementary material}
Sample codes can be found at
\url{https://github.com/mahsaashouri/Traffic-Anomaly}.

\begin{appendix}
\section{Additional data exploration} \label{app:ACFPACF}

\begin{figure}[!htp]
		\centering
		\subfloat[Northern region, Zhongli runway - Hukou, south direction, car]{\label{N2971S31-ACF}\includegraphics[width=0.75\textwidth]{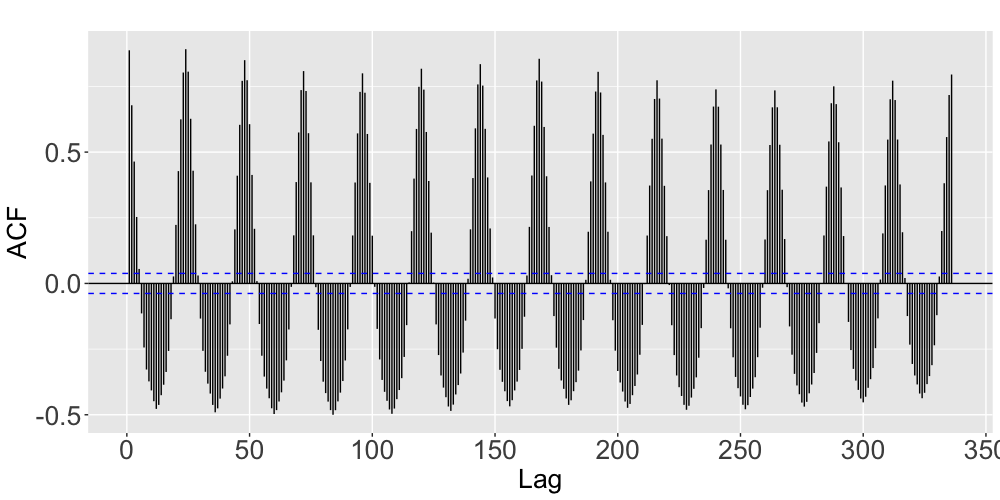}}\\
		\centering
		\subfloat[Central region, Xiangshan - Xibin, south direction, small truck]{\label{C2443S32-ACF}\includegraphics[width=0.75\textwidth]{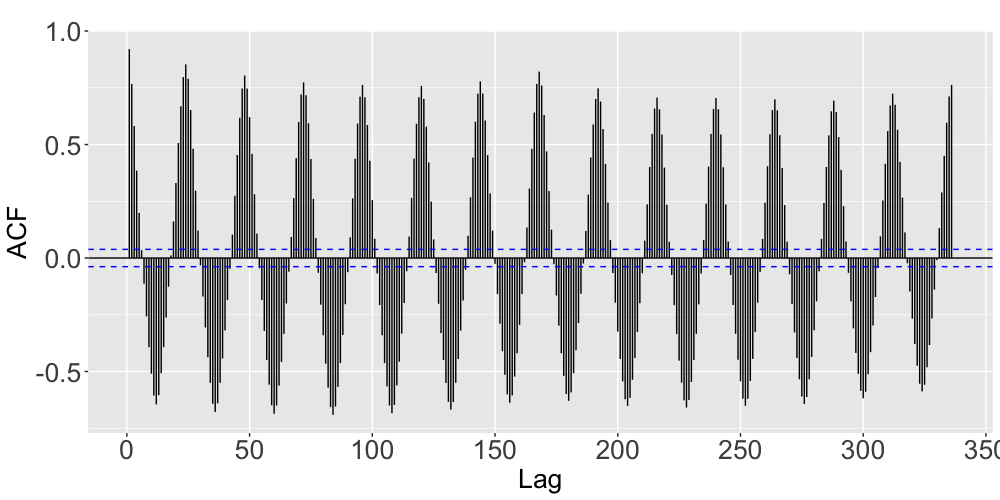}}\\
		\centering
		\subfloat[Southern region, Yanchao SIC - Tianliao, north direction, big truck]{\label{S2693N42-ACF}\includegraphics[width=0.75\textwidth]{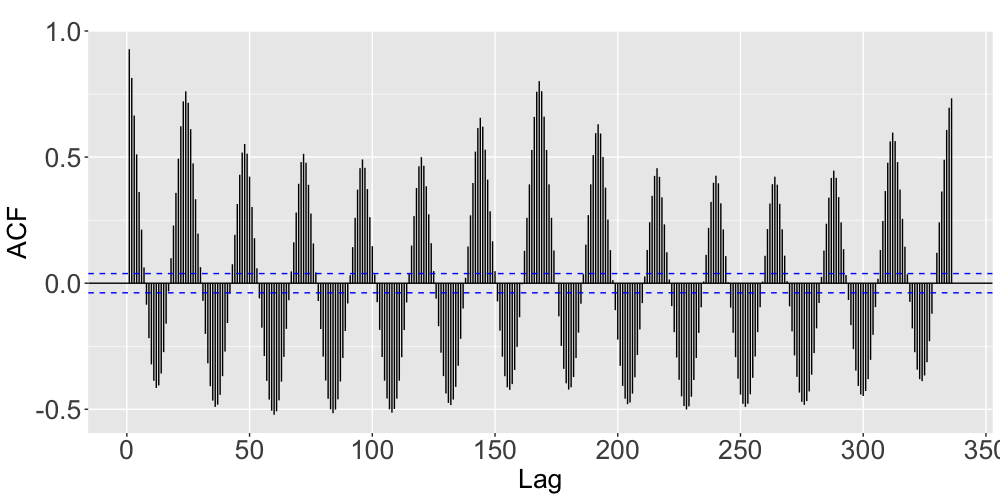}}
		\caption{
			ACF and PACF plots on examples of Taiwanese highway hourly time series (from 2021-04-23 23:00:00 to 2021-04-30 22:00:00) in three regions for different stations, traffic directions, and vehicle types. 
		}
		\label{fig:ACF-example-Taiwan-traffic-series}
	\end{figure}

\begin{figure}[!pht]
		\centering
  		\subfloat[Northern region, Zhongli runway - Hukou, south direction, car]{\label{N2971S31-PACF}\includegraphics[width=0.75\textwidth]{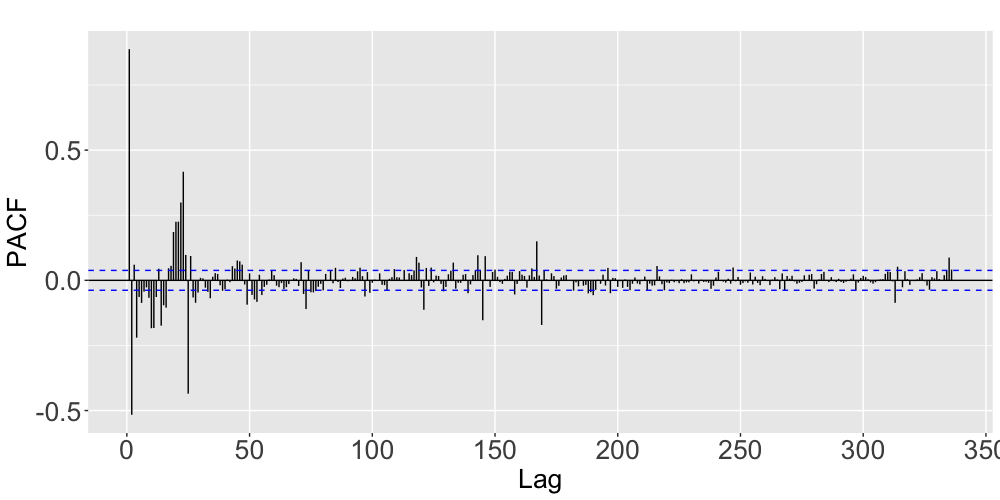}}\\
		\centering
  \subfloat[Central region, Xiangshan - Xibin, south direction, small truck]{\label{C2443S32-PACF}\includegraphics[width=0.75\textwidth]{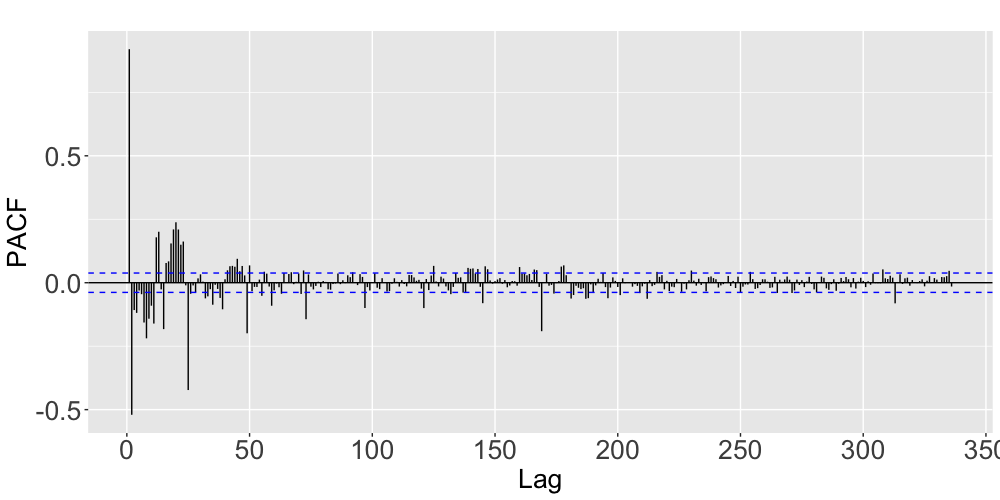}}\\
  \centering
  \subfloat[Southern region, Yanchao SIC - Tianliao, north direction, big truck]{\label{S2693N42-PACF}\includegraphics[width=0.75\textwidth]{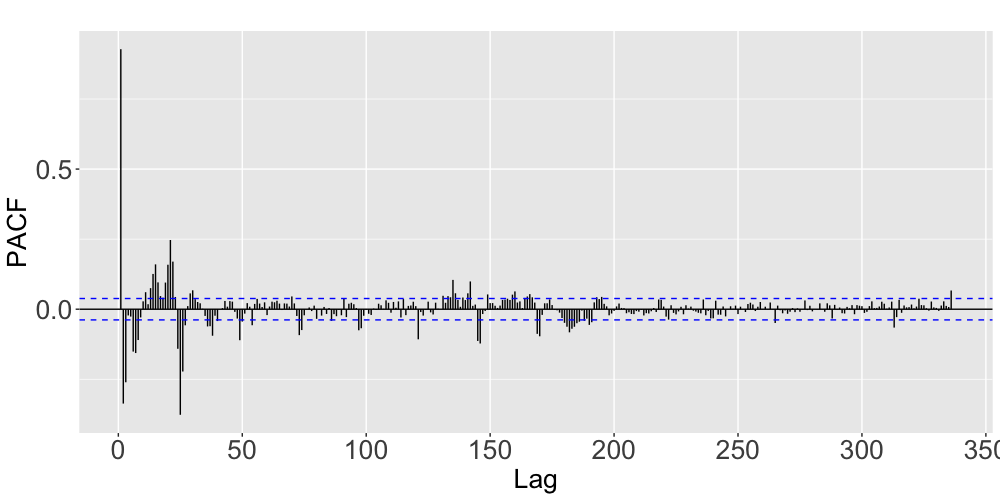}}
		\caption{
		PACF plots on examples of Taiwanese highway hourly time series (from 2021-04-23 23:00:00 to 2021-04-30 22:00:00) in three regions for different stations, traffic directions, and vehicle types. 
		}
		\label{fig:PACF-example-Taiwan-traffic-series}
	\end{figure}

Figures \ref{fig:ACF-example-Taiwan-traffic-series} and \ref{fig:PACF-example-Taiwan-traffic-series} display autocorrelation functions (ACF) and partial autocorrelation functions (PACF), respectively, for three examples of Taiwanese highway hourly time series (from 2021-04-16 23:00:00 to 2021-04-30 22:00:00 - last two weeks) in three regions for different stations, traffic directions, and vehicle types (the same series plotted in Figure \ref{fig:example-Taiwan-traffic-series}). 
Figure \ref{fig:ACF-example-Taiwan-traffic-series} shows a clear seasonal pattern in the data in all three series, corresponding to lag 24, i.e.~to a period of 1 day. 
This suggests that incorporating daily patterns into the data modeling process is appropriate (i.e.,~using Fourier terms with a period of $24$ hours in the OLS model). 
Figure \ref{S2693N42-ACF} also shows an additional seasonal pattern with a period of one week (i.e.,~with lag $168=7\times24$ hours). This seasonality appears to be present, although much less pronounced, also in the other two series (Figures \ref{N2971S31-ACF} and \ref{C2443S32-ACF}), emphasizing the necessity of incorporating also a weekly periodicity when modeling the data (i.e.,~using Fourier terms with a period of $7\times24$ hours in the OLS model). 
Figure \ref{fig:PACF-example-Taiwan-traffic-series} shows a significant spike at lag 1, whereas the lowest significant values occur at lags 2 and 24. These results suggest that utilizing lags 1 and 24 to capture autocorrelation in the OLS model is suitable. 

\section{Additional analysis of anomalies} \label{app:extraTableFigure}
\begin{sidewaysfigure}[ht]
		\centering
		\begin{tikzpicture}
			\node[scale=0.35]
			{ \scalebox{1}[1]{\includegraphics[scale=0.7]{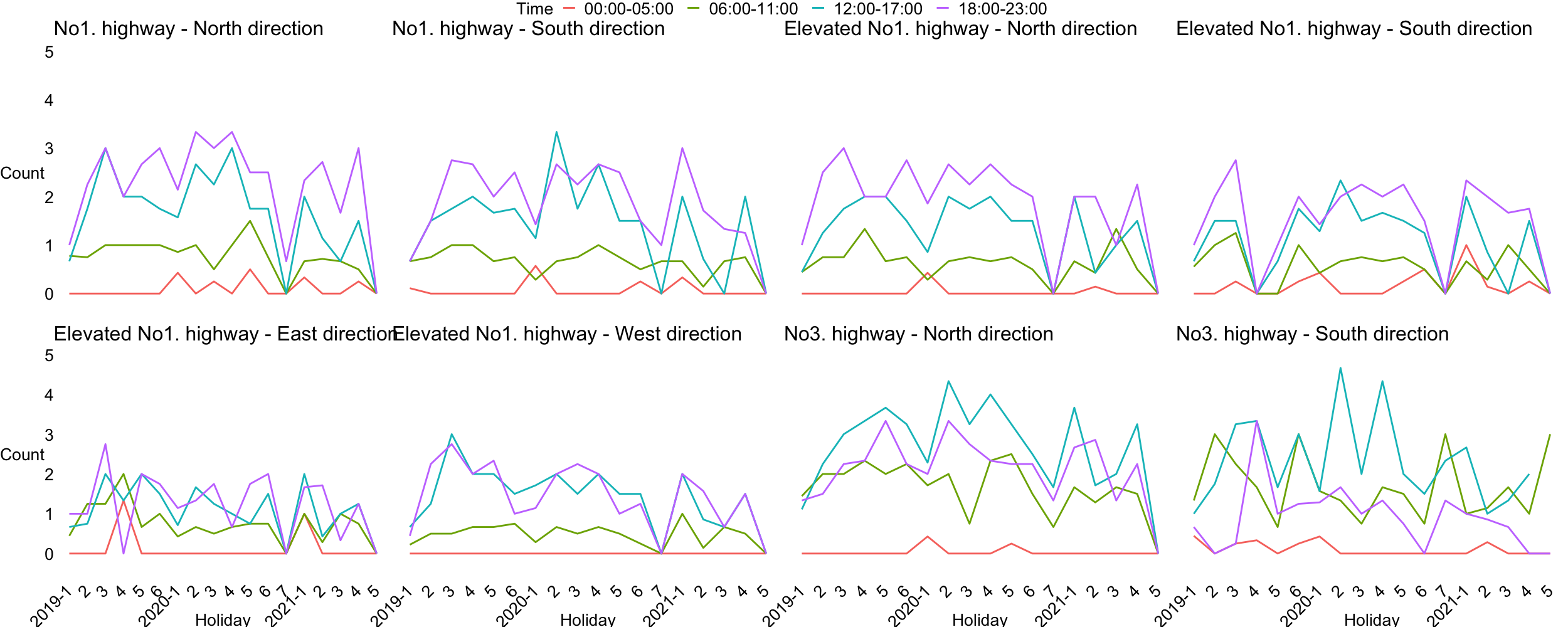}} };  
		\end{tikzpicture}
		\caption{
			Number of detected anomalies per day (divided by holiday length) in the 18 long holidays in highways No.1, Elevated No.1, and No.3 in the north, south, east, and west directions, divided into four time intervals: 00:00-05-00 (1), 06:00-11:00 (2), 12:00-17:00 (3), and 18:00-23:00 (4).}
		\label{fig:anomalies-highway-by-direction}
	\end{sidewaysfigure}

Tables \ref{tab:anomalies-highway-by-directionNS} and \ref{tab:anomalies-highway-by-directionEW} show the anomaly counts per day (divided by holiday length) of each holiday in each of the three highways and in four directions. These tables comprise four time intervals within a day, namely: 00:00 to 05:00, 06:00 to 11:00, 12:00 to 17:00, and 18:00 to 23:00.
The same information is depicted in Figure \ref{fig:anomalies-highway-by-direction}.

 \begin{table}[!tb]
\footnotesize
\caption{\label{tab:anomalies-highway-by-directionNS} 
			The number of detected anomalies per day (divided by holiday length) in the 18 long holidays in highways No.1, Elevated No.1, and No.3 in the north and south directions, divided into four time intervals: 00:00-05-00 (1), 06:00-11:00 (2), 12:00-17:00 (3), and 18:00-23:00 (4). Cell background colors are proportional to the number of anomalies detected (white for 0-1.5 anomalies, light gray for 1.5-3 anomalies, gray for 3-4.5 anomalies, and dark gray for 4.5 or more anomalies).
		}
    \centering
    \begin{tabular}{lrrrrrrrrrrrr}
     \toprule
			\multicolumn{1}{c}{} & \multicolumn{12}{c}{North direction} \\
			\multicolumn{1}{c}{} & \multicolumn{4}{c}{No.1} & \multicolumn{4}{c}{Elevated No.1} & \multicolumn{4}{c}{No.3}\\
			\cmidrule(l{3pt}r{3pt}){2-5} \cmidrule(l{3pt}r{3pt}){6-9} \cmidrule(l{3pt}r{3pt}){10-13} 
			Holiday &1 & 2 & 3 & 4 &1 & 2 & 3 & 4 &1 & 2 & 3 & 4 \\
			\midrule
			2019&&&&&&&&&&&&\\
			02-02 to 02-10 &0.0 &0.7 &0.6 &1.0& 0.0 &0.4& 0.4 &1.0& 0.0 &1.4& 1.1& 1.3\\
			02-28 to 03-03 &0.0& 0.7& \cellcolor{gray!15}1.7& \cellcolor{gray!15}2.2& 0.0& 0.7& 1.2& \cellcolor{gray!15}2.5& 0.0& \cellcolor{gray!15}2.0& \cellcolor{gray!15}2.2& 1.5\\
			04-04 to 04-07 
             &0.0 &1.0 &\cellcolor{gray!15}3.0 &\cellcolor{gray!15}3.0 &0.0 &0.75 &\cellcolor{gray!15}1.7& \cellcolor{gray!15}3.0 &0.0& \cellcolor{gray!15}2.0 &\cellcolor{gray!15}3.0& \cellcolor{gray!15}2.2\\
			06-07 to 06-09 
             & 0.0 &1.0 &\cellcolor{gray!15}2.0 &\cellcolor{gray!15}2.0& 0.0& 1.3& \cellcolor{gray!15}2.0& \cellcolor{gray!15}2.0 &0.0& \cellcolor{gray!15}2.3& \cellcolor{gray!40}3.3 &\cellcolor{gray!15}2.3 \\
			09-13 to 09-15 
           &0.0& 1.0 &\cellcolor{gray!15}2.0 &\cellcolor{gray!15}2.6 &0.0 &0.6 &\cellcolor{gray!15}2.0 &\cellcolor{gray!15}2.0 &0.0& \cellcolor{gray!15}2.0& \cellcolor{gray!40}3.6 &\cellcolor{gray!40}3.3\\
			10-10 to 10-13 
            &0.0 &1.0&\cellcolor{gray!15} 1.7 &\cellcolor{gray!15}3.0 &0.0 &0.7& 1.5 &\cellcolor{gray!15}2.7& 0.0& \cellcolor{gray!15}2.2& \cellcolor{gray!40}3.2& \cellcolor{gray!15}2.2\\
			2020&&&&&&&&&&&&\\
   			01-23 to 01-29 
             &0.4 &0.8& 1.5 &\cellcolor{gray!15}2.1& 0.4 &0.2& 0.8 &\cellcolor{gray!15}1.8& 0.4 &\cellcolor{gray!15}1.7 &\cellcolor{gray!15}2.2 &\cellcolor{gray!15}2.0\\
			02-28 to 03-01 
             &0.0& 1.0 &\cellcolor{gray!15}2.6& \cellcolor{gray!40}3.3 &0.0 &0.6 &\cellcolor{gray!15}2.0 &\cellcolor{gray!15}2.6& 0.0& \cellcolor{gray!15}2.0 &\cellcolor{gray!40}4.3& \cellcolor{gray!40}3.3\\
			04-02 to 04-05 
            &0.2& 0.5& \cellcolor{gray!15}2.2 &\cellcolor{gray!15}3.0& 0.0 &0.7& \cellcolor{gray!15}1.7 &\cellcolor{gray!15}2.2& 0.0 &0.7 &\cellcolor{gray!40}3.2& \cellcolor{gray!15}2.7 \\
		05-01 to 05-03 
            &0.0& 1.0& \cellcolor{gray!15}3.0 &\cellcolor{gray!40}3.3& 0.0& 0.6 &\cellcolor{gray!15}2.0 &\cellcolor{gray!15}2.6 &0.0 &\cellcolor{gray!15}2.3 &\cellcolor{gray!40}4.0 &\cellcolor{gray!15}2.3 \\
			06-25 to 06-28  
           &0.5 &1.5 &\cellcolor{gray!15}1.7 &\cellcolor{gray!15}2.5 &0.2 &0.7 &1.5& \cellcolor{gray!15}2.2 &0.0& \cellcolor{gray!15}2.2&\cellcolor{gray!40} 3.2& \cellcolor{gray!15}2.2\\
			10-01 to 10-04 
            &0.0 &0.7 &\cellcolor{gray!15}1.7 &\cellcolor{gray!15}2.5& 0.0& 0.5& 1.5& \cellcolor{gray!15}2.0& 0.0& 1.5& \cellcolor{gray!15}2.5& \cellcolor{gray!15}2.2\\
		    10-09 to 10-11 
            &0.0 &0.0& 0.0 &0.6& 0.0 &0.0& 0.0 &0.0& 0.0& 0.6& \cellcolor{gray!15}1.6& 1.3\\
            2021&&&&&&&&&&&&\\
		01-01 to 01-03  
            &0.3 &0.6& \cellcolor{gray!15}2.0 &\cellcolor{gray!15}2.3 &0.0& 0.6& \cellcolor{gray!15}2.0& \cellcolor{gray!15}2.0& 0.0 &\cellcolor{gray!15}1.6& \cellcolor{gray!40}3.6& \cellcolor{gray!15}2.6\\
			02-10 to 02-16  
           &0.0 &0.7& 1.4 &\cellcolor{gray!15}2.7& 0.0 &0.4 &0.4 &\cellcolor{gray!15}2.0& 0.1& 1.2& \cellcolor{gray!15}1.7 &\cellcolor{gray!15}2.8\\
			02-27 to 03-01 
           &0.0 &0.6 &0.6 &\cellcolor{gray!15}1.6 &0.0 &1.3& 1.0 &1.0& 0.0& \cellcolor{gray!15}1.6 &\cellcolor{gray!15}2.0& 1.3\\
			04-02 to 04-05 
          &0.2 &0.5 &1.5 &\cellcolor{gray!15}3.0 &0.0 &0.5& 1.5 &\cellcolor{gray!15}2.2 &0.0& 1.5& \cellcolor{gray!40}3.2& \cellcolor{gray!15}2.2\\
		04-30 to 04-30 
           &0.0&0.0&0.0&0.0&0.0&0.0&0.0&0.0&0.0&0.0&0.0&0.0\\
           \midrule
           \multicolumn{1}{c}{} & \multicolumn{12}{c}{South direction} \\
			\multicolumn{1}{c}{} & \multicolumn{4}{c}{No.1} & \multicolumn{4}{c}{Elevated No.1} & \multicolumn{4}{c}{No.3}\\
			\cmidrule(l{3pt}r{3pt}){2-5} \cmidrule(l{3pt}r{3pt}){6-9} \cmidrule(l{3pt}r{3pt}){10-13} 
			Holiday &1 & 2 & 3 & 4 &1 & 2 & 3 & 4 &1 & 2 & 3 & 4 \\
			\midrule
			2019&&&&&&&&&&&&\\
			02-02 to 02-10 & 0.1& 0.6& 0.6& 0.6& 0.0& 0.5& 0.6& 1.0& 0.4& 1.3& 1.0& 0.6\\
			02-28 to 03-03 & 0.0& 0.75& 1.5& 1.5& 0.0& 1.0& 1.5& \cellcolor{gray!15}2.0& 0.0& \cellcolor{gray!15}3.0& \cellcolor{gray!15}1.7& 0.0\\
			04-04 to 04-07 
             & 0.0& 1.0& \cellcolor{gray!15}1.7 &\cellcolor{gray!15}2.7& 0.2 &1.2 &1.5 &\cellcolor{gray!15}2.7& 0.2& \cellcolor{gray!15}2.2& \cellcolor{gray!40}3.2& 0.2\\
			06-07 to 06-09 
            &0.0 &1.0& \cellcolor{gray!15}2.0 &\cellcolor{gray!15}2.6 &0.0 &0.0 &0.0 &0.0& 0.3& \cellcolor{gray!15}1.6& \cellcolor{gray!40}3.3& \cellcolor{gray!40}3.3\\    
			09-13 to 09-15 
             & 0.0 &0.6& \cellcolor{gray!15}1.6& \cellcolor{gray!15}2.0 &0.0 &0.0 &0.6 &1.0 &0.0 &0.6 &\cellcolor{gray!15}1.6& 1.0\\    
			10-10 to 10-13 
             & 0.0& 0.7& \cellcolor{gray!15}1.7& \cellcolor{gray!15}2.5& 0.2 &1.0& \cellcolor{gray!15}1.7& \cellcolor{gray!15}2.0& 0.2& \cellcolor{gray!15}3.0& \cellcolor{gray!15}3.0 &1.2 \\ 
			2020&&&&&&&&&&&&\\
   			01-23 to 01-29 
             & 0.5& 0.2 &1.1& 1.4 &0.4& 0.4& 1.2& 1.4& 0.4& 1.5 &1.5& 1.2\\
			02-28 to 03-01 
             & 0.0 &0.6& \cellcolor{gray!40}3.3 &\cellcolor{gray!15}2.6 &0.0 &0.6 &\cellcolor{gray!15}2.3& \cellcolor{gray!15}2.0 &0.0 &1.3& \cellcolor{gray!60}4.6& \cellcolor{gray!15}1.6\\
			04-02 to 04-05 
            &0.0& 0.7& \cellcolor{gray!15}1.7 &\cellcolor{gray!15}2.2& 0.0 &0.7& 1.5& \cellcolor{gray!15}2.2& 0.0& 0.7& \cellcolor{gray!15}2.0& 1.0\\
		05-01 to 05-03 
             &0.0 &1.0 &\cellcolor{gray!15}2.6 &\cellcolor{gray!15}2.6& 0.0 &0.6 &\cellcolor{gray!15}1.6 &\cellcolor{gray!15}2.0 &0.0 &\cellcolor{gray!15}1.6 &\cellcolor{gray!40}4.3 &1.3 \\
			06-25 to 06-28  
            & 0.0& 0.7& 1.5 &\cellcolor{gray!15}2.5 &0.2& 0.7 &1.5& \cellcolor{gray!15}2.2& 0.0& 1.5& \cellcolor{gray!15}2.0& 0.7\\
			10-01 to 10-04 
           &0.2& 0.5 &1.5& 1.5& 0.5& 0.5& 1.2 &1.5 &0.0& 0.7& 1.5& 0.0\\
		    10-09 to 10-11 
           & 0.0& 0.6 &0.0& 1.0& 0.0& 0.0 &0.0& 0.0& 0.0& \cellcolor{gray!15}3.0 &\cellcolor{gray!15}2.3& 1.3\\
            2021&&&&&&&&&&&&\\
		01-01 to 01-03  
            & 0.3& 0.6 &\cellcolor{gray!15}2.0& \cellcolor{gray!15}3.0& 1.0 &0.6& \cellcolor{gray!15}2.0 &\cellcolor{gray!15}2.3 &0.0 &1.0 &\cellcolor{gray!40}3.6& 1.0\\
			02-10 to 02-16  
           & 0.0& 0.1 &0.7 &\cellcolor{gray!15}1.7& 0.1 &0.2& 0.8 &\cellcolor{gray!15}2.0 &0.2& 1.1& 1.0& 0.8\\
			02-27 to 03-01 
           &0.0& 0.6& 0.0& 1.3& 0.0 &1.0& 0.0& \cellcolor{gray!15}1.6 &0.0& \cellcolor{gray!15}1.6& 1.3 &0.6 \\
			04-02 to 04-05 
           & 0.0& 0.7& \cellcolor{gray!15}2.0 &1.2& 0.2& 0.5& 1.5& \cellcolor{gray!15}1.7& 0.0& 1.0& \cellcolor{gray!15}2.0& 0.0\\
		04-30 to 04-30 
           &0.0&0.0&0.0&0.0&0.0&0.0&0.0&0.0&0.0&\cellcolor{gray!15}3.0&\cellcolor{gray!60}6.0&0.0\\
			\bottomrule
		\end{tabular}
\end{table}

 \begin{table}[!tb]
\footnotesize
\caption{\label{tab:anomalies-highway-by-directionEW} 
			The number of detected anomalies per day (divided by holiday length) in the 18 long holidays in highway Elevated No.1 in the east and west directions, divided into four time intervals: 00:00-05-00 (1), 06:00-11:00 (2), 12:00-17:00 (3), and 18:00-23:00 (4). Cell background colors are proportional to the number of anomalies detected (white for 0-1.5 anomalies, light gray for 1.5-3 anomalies, gray for 3-4.5 anomalies, and dark gray for 4.5 or more anomalies).
		}
		\centering
		\begin{tabular}[ht]{lrrrrrrrr}
			\toprule
			\multicolumn{1}{c}{} & \multicolumn{4}{c}{East direction} &  \multicolumn{4}{c}{West direction}\\
			\cmidrule(l{3pt}r{3pt}){2-5} \cmidrule(l{3pt}r{3pt}){6-9} 
			\multicolumn{1}{c}{} & \multicolumn{4}{c}{Elevated No.1} & \multicolumn{4}{c}{Elevated No.1}\\
			\cmidrule(l{3pt}r{3pt}){2-5} \cmidrule(l{3pt}r{3pt}){6-9}  
			Holiday &1 & 2 & 3 & 4 &1 & 2 & 3 & 4\\
			\midrule
			2019&&&&&&&&\\
			02-02 to 02-10  & 0.0& 0.4& 0.6& 1.0& 0.0& 0.2& 0.6 &0.4\\
			02-28 to 03-03& 0.0& 1.2& 0.7& 1.0& 0.0& 0.5& 1.2& \cellcolor{gray!15}2.2\\
			04-04 to 04-07 
             & 0.0& 1.2 &\cellcolor{gray!15}2.0& \cellcolor{gray!15}2.7 &0.0 &0.5&\cellcolor{gray!15} 3.0 &\cellcolor{gray!15}2.7\\
			06-07 to 06-09 
           &1.3& \cellcolor{gray!15}2.0& 1.3 &0.0& 0.0 &0.6&\cellcolor{gray!15} 2.0 &\cellcolor{gray!15} 2.0\\
			09-13 to 09-15 
           & 0.0& 0.6 &\cellcolor{gray!15}2.0 &\cellcolor{gray!15}2.0 &0.0 &0.6 &\cellcolor{gray!15}2.0 &\cellcolor{gray!15}2.3\\
			10-10 to 10-13 
          &0.0& 1.0& 1.5& \cellcolor{gray!15}1.7& 0.0& 0.7& 1.5 &1.0\\
			2020&&&&&&&&\\
   			01-23 to 01-29 
             & 0.0& 0.4 &0.7& 1.1& 0.0 &0.2 &\cellcolor{gray!15}1.7& 1.1\\
			02-28 to 03-01 
            & 0.0& 0.6& \cellcolor{gray!15}1.6& 1.3& 0.0& 0.6 &\cellcolor{gray!15}2.0& \cellcolor{gray!15}2.0\\
			04-02 to 04-05 
            &0.0& 0.5& 1.2 &\cellcolor{gray!15}1.7& 0.0& 0.5 &1.5& \cellcolor{gray!15}2.2\\
		05-01 to 05-03 
           &0.0 &0.6 &1.0 &0.6 &0.0 &0.6 &\cellcolor{gray!15}2.0& \cellcolor{gray!15}2.0\\
			06-25 to 06-28  
           &0.0 &0.7& 0.7& \cellcolor{gray!15}1.7& 0.0 &0.5 &1.5 &1.0\\
			10-01 to 10-04 
           & 0.0& 0.7 &1.5 &\cellcolor{gray!15}2.0 &0.0 &0.2& 1.5& 1.2\\
		    10-09 to 10-11 
             &0.0& 0.0 &0.0& 0.0& 0.0& 0.0 &0.0 &0.0\\
            2021&&&&&&&&\\
		01-01 to 01-03  
            &1.0& 1.0& \cellcolor{gray!15}2.0& \cellcolor{gray!15}1.6 &0.0 &1.0 &\cellcolor{gray!15}2.0& \cellcolor{gray!15}2.0\\
			02-10 to 02-16  
           & 0.0& 0.2& 0.4& \cellcolor{gray!15}1.7& 0.0& 0.1& 0.8 &1.5\\
			02-27 to 03-01 
           &0.0 &1.0 &1.0& 0.3& 0.0& 0.6&0.6 &0.6\\
			04-02 to 04-05 
          & 0.0& 0.7& 1.2 &1.2 &0.0& 0.5 &1.8& 1.5\\
		04-30 to 04-30 
           &0.0&0.0&0.0&0.0&0.0&0.0&0.0&0.0\\
			\bottomrule
		\end{tabular}
\end{table}
	
\end{appendix}


\bibliography{bibliography}       


\begin{thebibliography}{29}
\ifx \bisbn   \undefined \def \bisbn  #1{ISBN #1}\fi
\ifx \binits  \undefined \def \binits#1{#1}\fi
\ifx \bauthor  \undefined \def \bauthor#1{#1}\fi
\ifx \batitle  \undefined \def \batitle#1{#1}\fi
\ifx \bjtitle  \undefined \def \bjtitle#1{#1}\fi
\ifx \bvolume  \undefined \def \bvolume#1{\textbf{#1}}\fi
\ifx \byear  \undefined \def \byear#1{#1}\fi
\ifx \bissue  \undefined \def \bissue#1{#1}\fi
\ifx \bfpage  \undefined \def \bfpage#1{#1}\fi
\ifx \blpage  \undefined \def \blpage #1{#1}\fi
\ifx \burl  \undefined \def \burl#1{\textsf{#1}}\fi
\ifx \doiurl  \undefined \def \doiurl#1{\url{https://doi.org/#1}}\fi
\ifx \betal  \undefined \def \betal{\textit{et al.}}\fi
\ifx \binstitute  \undefined \def \binstitute#1{#1}\fi
\ifx \binstitutionaled  \undefined \def \binstitutionaled#1{#1}\fi
\ifx \bctitle  \undefined \def \bctitle#1{#1}\fi
\ifx \beditor  \undefined \def \beditor#1{#1}\fi
\ifx \bpublisher  \undefined \def \bpublisher#1{#1}\fi
\ifx \bbtitle  \undefined \def \bbtitle#1{#1}\fi
\ifx \bedition  \undefined \def \bedition#1{#1}\fi
\ifx \bseriesno  \undefined \def \bseriesno#1{#1}\fi
\ifx \blocation  \undefined \def \blocation#1{#1}\fi
\ifx \bsertitle  \undefined \def \bsertitle#1{#1}\fi
\ifx \bsnm \undefined \def \bsnm#1{#1}\fi
\ifx \bsuffix \undefined \def \bsuffix#1{#1}\fi
\ifx \bparticle \undefined \def \bparticle#1{#1}\fi
\ifx \barticle \undefined \def \barticle#1{#1}\fi
\bibcommenthead
\ifx \bconfdate \undefined \def \bconfdate #1{#1}\fi
\ifx \botherref \undefined \def \botherref #1{#1}\fi
\ifx \url \undefined \def \url#1{\textsf{#1}}\fi
\ifx \bchapter \undefined \def \bchapter#1{#1}\fi
\ifx \bbook \undefined \def \bbook#1{#1}\fi
\ifx \bcomment \undefined \def \bcomment#1{#1}\fi
\ifx \oauthor \undefined \def \oauthor#1{#1}\fi
\ifx \citeauthoryear \undefined \def \citeauthoryear#1{#1}\fi
\ifx \endbibitem  \undefined \def \endbibitem {}\fi
\ifx \bconflocation  \undefined \def \bconflocation#1{#1}\fi
\ifx \arxivurl  \undefined \def \arxivurl#1{\textsf{#1}}\fi
\csname PreBibitemsHook\endcsname

\bibitem[\protect\citeauthoryear{Chu et~al.}{2015}]{chu2015road}
\begin{bchapter}
\bauthor{\bsnm{Chu}, \binits{C.}},
\bauthor{\bsnm{Hu}, \binits{S.-R.}},
\bauthor{\bsnm{Chiang}, \binits{C.}},
\bauthor{\bsnm{Lu}, \binits{Y.}}:
\bctitle{Road space rationing policies for freeway holiday congestion
  management in taiwan-a simulation approach}.
In: \bbtitle{20th International Conference of Hong Kong Society for
  Transportation Studies: Urban Transport Analytics, HKSTS 2015},
pp. \bfpage{287}--\blpage{293}
(\byear{2015}).
\bcomment{Hong Kong Society for Transportation Studies Limited}
\end{bchapter}
\endbibitem

\bibitem[\protect\citeauthoryear{Darban et~al.}{2022}]{darban2022deep}
\begin{botherref}
\oauthor{\bsnm{Darban}, \binits{Z.Z.}},
\oauthor{\bsnm{Webb}, \binits{G.I.}},
\oauthor{\bsnm{Pan}, \binits{S.}},
\oauthor{\bsnm{Aggarwal}, \binits{C.C.}},
\oauthor{\bsnm{Salehi}, \binits{M.}}:
Deep learning for time series anomaly detection: A survey.
arXiv preprint arXiv:2211.05244
(2022)
\end{botherref}
\endbibitem

\bibitem[\protect\citeauthoryear{Mondal and Rehena}{2020}]{mondal2020road}
\begin{barticle}
\bauthor{\bsnm{Mondal}, \binits{M.A.}},
\bauthor{\bsnm{Rehena}, \binits{Z.}}:
\batitle{Road traffic outlier detection technique based on linear regression}.
\bjtitle{Procedia Computer Science}
\bvolume{171},
\bfpage{2547}--\blpage{2555}
(\byear{2020})
\end{barticle}
\endbibitem

\bibitem[\protect\citeauthoryear{Tang and Gao}{2005}]{tang2005traffic}
\begin{barticle}
\bauthor{\bsnm{Tang}, \binits{S.}},
\bauthor{\bsnm{Gao}, \binits{H.}}:
\batitle{Traffic-incident detection-algorithm based on nonparametric
  regression}.
\bjtitle{IEEE Transactions on Intelligent Transportation Systems}
\bvolume{6}(\bissue{1}),
\bfpage{38}--\blpage{42}
(\byear{2005})
\end{barticle}
\endbibitem

\bibitem[\protect\citeauthoryear{Djenouri et~al.}{2019}]{djenouri2019survey}
\begin{barticle}
\bauthor{\bsnm{Djenouri}, \binits{Y.}},
\bauthor{\bsnm{Belhadi}, \binits{A.}},
\bauthor{\bsnm{Lin}, \binits{J.C.-W.}},
\bauthor{\bsnm{Djenouri}, \binits{D.}},
\bauthor{\bsnm{Cano}, \binits{A.}}:
\batitle{A survey on urban traffic anomalies detection algorithms}.
\bjtitle{IEEE Access}
\bvolume{7},
\bfpage{12192}--\blpage{12205}
(\byear{2019})
\end{barticle}
\endbibitem

\bibitem[\protect\citeauthoryear{Bawaneh and Simon}{2019}]{bawaneh2019anomaly}
\begin{bchapter}
\bauthor{\bsnm{Bawaneh}, \binits{M.}},
\bauthor{\bsnm{Simon}, \binits{V.}}:
\bctitle{Anomaly detection in smart city traffic based on time series
  analysis}.
In: \bbtitle{2019 International Conference on Software, Telecommunications and
  Computer Networks (SoftCOM)},
pp. \bfpage{1}--\blpage{6}
(\byear{2019}).
\bcomment{IEEE}
\end{bchapter}
\endbibitem

\bibitem[\protect\citeauthoryear{Huang et~al.}{2018}]{huang2018traffic}
\begin{botherref}
\oauthor{\bsnm{Huang}, \binits{T.}},
\oauthor{\bsnm{Liu}, \binits{C.}},
\oauthor{\bsnm{Sharma}, \binits{A.}},
\oauthor{\bsnm{Sarkar}, \binits{S.}}:
Traffic system anomaly detection using spatiotemporal pattern networks.
International Journal of Prognostics and Health Management
\textbf{9}(1)
(2018)
\end{botherref}
\endbibitem

\bibitem[\protect\citeauthoryear{Zhang et~al.}{2016}]{zhang2016spatial}
\begin{barticle}
\bauthor{\bsnm{Zhang}, \binits{Z.}},
\bauthor{\bsnm{He}, \binits{Q.}},
\bauthor{\bsnm{Tong}, \binits{H.}},
\bauthor{\bsnm{Gou}, \binits{J.}},
\bauthor{\bsnm{Li}, \binits{X.}}:
\batitle{Spatial-temporal traffic flow pattern identification and anomaly
  detection with dictionary-based compression theory in a large-scale urban
  network}.
\bjtitle{Transportation Research Part C: Emerging Technologies}
\bvolume{71},
\bfpage{284}--\blpage{302}
(\byear{2016})
\end{barticle}
\endbibitem

\bibitem[\protect\citeauthoryear{Riveiro et~al.}{2017}]{riveiro2017anomaly}
\begin{barticle}
\bauthor{\bsnm{Riveiro}, \binits{M.}},
\bauthor{\bsnm{Lebram}, \binits{M.}},
\bauthor{\bsnm{Elmer}, \binits{M.}}:
\batitle{Anomaly detection for road traffic: A visual analytics framework}.
\bjtitle{IEEE Transactions on Intelligent Transportation Systems}
\bvolume{18}(\bissue{8}),
\bfpage{2260}--\blpage{2270}
(\byear{2017})
\end{barticle}
\endbibitem

\bibitem[\protect\citeauthoryear{Aboah}{2021}]{aboah2021vision}
\begin{bchapter}
\bauthor{\bsnm{Aboah}, \binits{A.}}:
\bctitle{A vision-based system for traffic anomaly detection using deep
  learning and decision trees}.
In: \bbtitle{Proceedings of the IEEE/CVF Conference on Computer Vision and
  Pattern Recognition},
pp. \bfpage{4207}--\blpage{4212}
(\byear{2021})
\end{bchapter}
\endbibitem

\bibitem[\protect\citeauthoryear{Hou et~al.}{2013}]{hou2013detection}
\begin{barticle}
\bauthor{\bsnm{Hou}, \binits{D.}},
\bauthor{\bsnm{He}, \binits{H.}},
\bauthor{\bsnm{Huang}, \binits{P.}},
\bauthor{\bsnm{Zhang}, \binits{G.}},
\bauthor{\bsnm{Loaiciga}, \binits{H.}}:
\batitle{Detection of water-quality contamination events based on multi-sensor
  fusion using an extented dempster--shafer method}.
\bjtitle{Measurement Science and Technology}
\bvolume{24}(\bissue{5}),
\bfpage{055801}
(\byear{2013})
\end{barticle}
\endbibitem

\bibitem[\protect\citeauthoryear{Li et~al.}{2019}]{li2019detection}
\begin{barticle}
\bauthor{\bsnm{Li}, \binits{X.}},
\bauthor{\bsnm{Zhang}, \binits{T.}},
\bauthor{\bsnm{Liu}, \binits{Y.}}:
\batitle{Detection of voltage anomalies in spacecraft storage batteries based
  on a deep belief network}.
\bjtitle{Sensors}
\bvolume{19}(\bissue{21}),
\bfpage{4702}
(\byear{2019})
\end{barticle}
\endbibitem

\bibitem[\protect\citeauthoryear{Pang et~al.}{2018}]{pang2018optimize}
\begin{barticle}
\bauthor{\bsnm{Pang}, \binits{J.}},
\bauthor{\bsnm{Liu}, \binits{D.}},
\bauthor{\bsnm{Peng}, \binits{Y.}},
\bauthor{\bsnm{Peng}, \binits{X.}}:
\batitle{Optimize the coverage probability of prediction interval for anomaly
  detection of sensor-based monitoring series}.
\bjtitle{Sensors}
\bvolume{18}(\bissue{4}),
\bfpage{967}
(\byear{2018})
\end{barticle}
\endbibitem

\bibitem[\protect\citeauthoryear{Zhong et~al.}{2023}]{zhong2023survey}
\begin{botherref}
\oauthor{\bsnm{Zhong}, \binits{Z.}},
\oauthor{\bsnm{Fan}, \binits{Q.}},
\oauthor{\bsnm{Zhang}, \binits{J.}},
\oauthor{\bsnm{Ma}, \binits{M.}},
\oauthor{\bsnm{Zhang}, \binits{S.}},
\oauthor{\bsnm{Sun}, \binits{Y.}},
\oauthor{\bsnm{Lin}, \binits{Q.}},
\oauthor{\bsnm{Zhang}, \binits{Y.}},
\oauthor{\bsnm{Pei}, \binits{D.}}:
A survey of time series anomaly detection methods in the aiops domain.
arXiv preprint arXiv:2308.00393
(2023)
\end{botherref}
\endbibitem

\bibitem[\protect\citeauthoryear{Choffnes
  et~al.}{2010}]{choffnes2010crowdsourcing}
\begin{bchapter}
\bauthor{\bsnm{Choffnes}, \binits{D.R.}},
\bauthor{\bsnm{Bustamante}, \binits{F.E.}},
\bauthor{\bsnm{Ge}, \binits{Z.}}:
\bctitle{Crowdsourcing service-level network event monitoring}.
In: \bbtitle{Proceedings of the ACM SIGCOMM 2010 Conference},
pp. \bfpage{387}--\blpage{398}
(\byear{2010})
\end{bchapter}
\endbibitem

\bibitem[\protect\citeauthoryear{Zhang et~al.}{2005}]{zhang2005network}
\begin{bchapter}
\bauthor{\bsnm{Zhang}, \binits{Y.}},
\bauthor{\bsnm{Ge}, \binits{Z.}},
\bauthor{\bsnm{Greenberg}, \binits{A.}},
\bauthor{\bsnm{Roughan}, \binits{M.}}:
\bctitle{Network anomography}.
In: \bbtitle{Proceedings of the 5th ACM SIGCOMM Conference on Internet
  Measurement},
pp. \bfpage{30}--\blpage{30}
(\byear{2005})
\end{bchapter}
\endbibitem

\bibitem[\protect\citeauthoryear{Qiu et~al.}{2019}]{qiu2019short}
\begin{bchapter}
\bauthor{\bsnm{Qiu}, \binits{J.}},
\bauthor{\bsnm{Du}, \binits{Q.}},
\bauthor{\bsnm{Wang}, \binits{W.}},
\bauthor{\bsnm{Yin}, \binits{K.}},
\bauthor{\bsnm{Chen}, \binits{L.}}:
\bctitle{Short-term performance metrics forecasting for virtual machine to
  support anomaly detection using hybrid arima-wnn model}.
In: \bbtitle{2019 IEEE 43rd Annual Computer Software and Applications
  Conference (COMPSAC)},
vol. \bseriesno{2},
pp. \bfpage{330}--\blpage{335}
(\byear{2019}).
\bcomment{IEEE}
\end{bchapter}
\endbibitem

\bibitem[\protect\citeauthoryear{Siu et~al.}{2020}]{siu2020switching}
\begin{barticle}
\bauthor{\bsnm{Siu}, \binits{K.T.}},
\bauthor{\bsnm{Xu}, \binits{Y.}},
\bauthor{\bsnm{Tai}, \binits{Y.M.}},
\bauthor{\bsnm{Choi}, \binits{T.}},
\bauthor{\bsnm{Michael~Wong}, \binits{K.}},
\bauthor{\bsnm{To}, \binits{K.}}:
\batitle{Switching behavior and control policy of congestion: Examples from
  taiwan highway system}.
\bjtitle{New Mathematics and Natural Computation}
\bvolume{16}(\bissue{03}),
\bfpage{657}--\blpage{667}
(\byear{2020})
\end{barticle}
\endbibitem

\bibitem[\protect\citeauthoryear{Ashouri et~al.}{2021}]{ashouri2021fast}
\begin{botherref}
\oauthor{\bsnm{Ashouri}, \binits{M.}},
\oauthor{\bsnm{Hyndman}, \binits{R.J.}},
\oauthor{\bsnm{Shmueli}, \binits{G.}}:
Fast forecast reconciliation using linear models.
Journal of Computational and Graphical Statistics,
1--20
(2021)
\end{botherref}
\endbibitem

\bibitem[\protect\citeauthoryear{Hyndman and Athanasopoulos}{2018}]{fpp3}
\begin{bbook}
\bauthor{\bsnm{Hyndman}, \binits{R.J.}},
\bauthor{\bsnm{Athanasopoulos}, \binits{G.}}:
\bbtitle{Forecasting: Principles and Practice}.
\bpublisher{OTexts},
\blocation{Melbourne, Australia}
(\byear{2018}).
\burl{https://OTexts.org/fpp3}
\end{bbook}
\endbibitem

\bibitem[\protect\citeauthoryear{Kahn}{1998}]{kahn1998revisiting}
\begin{barticle}
\bauthor{\bsnm{Kahn}, \binits{K.B.}}:
\batitle{Revisiting top-down versus bottom-up forecasting}.
\bjtitle{The Journal of Business Forecasting}
\bvolume{17}(\bissue{2}),
\bfpage{14}
(\byear{1998})
\end{barticle}
\endbibitem

\bibitem[\protect\citeauthoryear{Gross and
  Sohl}{1990}]{gross1990disaggregation}
\begin{barticle}
\bauthor{\bsnm{Gross}, \binits{C.W.}},
\bauthor{\bsnm{Sohl}, \binits{J.E.}}:
\batitle{Disaggregation methods to expedite product line forecasting}.
\bjtitle{Journal of Forecasting}
\bvolume{9}(\bissue{3}),
\bfpage{233}--\blpage{254}
(\byear{1990})
\end{barticle}
\endbibitem

\bibitem[\protect\citeauthoryear{Hyndman et~al.}{2011}]{hyndman2011optimal}
\begin{barticle}
\bauthor{\bsnm{Hyndman}, \binits{R.J.}},
\bauthor{\bsnm{Ahmed}, \binits{R.A.}},
\bauthor{\bsnm{Athanasopoulos}, \binits{G.}},
\bauthor{\bsnm{Shang}, \binits{H.L.}}:
\batitle{Optimal combination forecasts for hierarchical time series}.
\bjtitle{Computational Statistics \& Data Analysis}
\bvolume{55}(\bissue{9}),
\bfpage{2579}--\blpage{2589}
(\byear{2011})
\end{barticle}
\endbibitem

\bibitem[\protect\citeauthoryear{Wickramasuriya
  et~al.}{2019}]{wickramasuriya2019optimal}
\begin{barticle}
\bauthor{\bsnm{Wickramasuriya}, \binits{S.L.}},
\bauthor{\bsnm{Athanasopoulos}, \binits{G.}},
\bauthor{\bsnm{Hyndman}, \binits{R.J.}}:
\batitle{Optimal forecast reconciliation for hierarchical and grouped time
  series through trace minimization}.
\bjtitle{Journal of the American Statistical Association}
\bvolume{114}(\bissue{526}),
\bfpage{804}--\blpage{819}
(\byear{2019})
\end{barticle}
\endbibitem

\bibitem[\protect\citeauthoryear{Li et~al.}{2019}]{li2019short}
\begin{barticle}
\bauthor{\bsnm{Li}, \binits{Z.}},
\bauthor{\bsnm{Zheng}, \binits{Z.}},
\bauthor{\bsnm{Washington}, \binits{S.}}:
\batitle{Short-term traffic flow forecasting: a component-wise gradient
  boosting approach with hierarchical reconciliation}.
\bjtitle{IEEE Transactions on Intelligent Transportation Systems}
\bvolume{21}(\bissue{12}),
\bfpage{5060}--\blpage{5072}
(\byear{2019})
\end{barticle}
\endbibitem

\bibitem[\protect\citeauthoryear{Ashouri et~al.}{2018}]{ashouri2018assessing}
\begin{barticle}
\bauthor{\bsnm{Ashouri}, \binits{M.}},
\bauthor{\bsnm{Cai}, \binits{K.}},
\bauthor{\bsnm{Lin}, \binits{F.}},
\bauthor{\bsnm{Shmueli}, \binits{G.}}:
\batitle{Assessing the value of an information system for developing predictive
  analytics: The case of forecasting school-level demand in taiwan}.
\bjtitle{Service Science}
\bvolume{10}(\bissue{1}),
\bfpage{58}--\blpage{75}
(\byear{2018})
\end{barticle}
\endbibitem

\bibitem[\protect\citeauthoryear{Ashouri et~al.}{2019}]{ashouri2019tree}
\begin{barticle}
\bauthor{\bsnm{Ashouri}, \binits{M.}},
\bauthor{\bsnm{Shmueli}, \binits{G.}},
\bauthor{\bsnm{Sin}, \binits{C.-Y.}}:
\batitle{Tree-based methods for clustering time series using domain-relevant
  attributes}.
\bjtitle{Journal of Business Analytics}
\bvolume{2}(\bissue{1}),
\bfpage{1}--\blpage{23}
(\byear{2019})
\end{barticle}
\endbibitem

\bibitem[\protect\citeauthoryear{Davison and
  Hinkley}{1997}]{davison1997bootstrap}
\begin{bbook}
\bauthor{\bsnm{Davison}, \binits{A.C.}},
\bauthor{\bsnm{Hinkley}, \binits{D.V.}}:
\bbtitle{Bootstrap Methods and Their Application}
vol. \bseriesno{1}.
\bpublisher{Cambridge university press}, \blocation{???}
(\byear{1997})
\end{bbook}
\endbibitem

\bibitem[\protect\citeauthoryear{Hyndman and
  Khandakar}{2008}]{hyndman2008automatic}
\begin{barticle}
\bauthor{\bsnm{Hyndman}, \binits{R.J.}},
\bauthor{\bsnm{Khandakar}, \binits{Y.}}:
\batitle{Automatic time series forecasting: the forecast package for r}.
\bjtitle{Journal of statistical software}
\bvolume{27},
\bfpage{1}--\blpage{22}
(\byear{2008})
\end{barticle}
\endbibitem

\end{thebibliography}

\end{document}